\documentclass[structabstract]{aa}
\usepackage{txfonts}

\begin{document}

\title{Kinetic temperature of massive star forming
molecular clumps measured with formaldehyde. II. The Large Magellanic Cloud}

\author{X. D. Tang\inst{1,2,3}
\and C. Henkel\inst{1,4}
\and C. -H. R. Chen\inst{1}
\and K. M. Menten\inst{1}
\and R. Indebetouw\inst{5,6}
\and X. W. Zheng\inst{7}
\and J. Esimbek\inst{2,3}
\and J. J. Zhou\inst{2,3}
\and Y. Yuan\inst{2,3}
\and D. L. Li\inst{2,3}
\and Y. X. He\inst{2,3}}

\titlerunning{Kinetic temperatures in molecular clump of the LMC}
\authorrunning{X. D. Tang et al.}

\institute{ Max-Planck-Institut f\"{u}r Radioastronomie, Auf dem H\"{u}gel 69, 53121 Bonn, Germany\\
\email{xdtang@mpifr-bonn.mpg.de}
\and Xinjiang Astronomical Observatory, Chinese Academy of Sciences, 830011 Urumqi, China\\
\and Key Laboratory of Radio Astronomy, Chinese Academy of Sciences, 830011 Urumqi, China\\
\and Astronomy Department, King Abdulaziz University, PO Box 80203, 21589 Jeddah, Saudi Arabia\\
\and National Radio Astronomy Observatory, 520 Edgemont Rd, Charlottesville, VA 22903, USA\\
\and University of Virginia, Charlottesville, VA 22903, USA\\
\and Department of Astronomy, Nanjing University, 210093 Nanjing, China\\}


\abstract
{The kinetic temperature of molecular clouds is a fundamental physical
parameter affecting star formation and the initial mass function.
The Large Magellanic Cloud (LMC), the closest star forming galaxy with low
metallicity, provides an ideal laboratory to study star formation in such an environment.}
{The classical dense
molecular gas thermometer NH$_3$ is rarely available in a low
metallicity environment because of photoionization and a lack of nitrogen atoms.
Our goal is to directly measure the gas kinetic temperature with
formaldehyde toward six star-forming regions in the LMC.}
{Three rotational transitions ($J$$_{\rm K_AK_C}$ = 3$_{03}$-2$_{02}$,
3$_{22}$-2$_{21}$, and 3$_{21}$-2$_{20}$) of para-H$_2$CO near 218 GHz were
observed with the Atacama Pathfinder
EXperiment (APEX) 12 m telescope toward six star forming regions
in the LMC. Those data are complemented by C$^{18}$O 2-1 spectra.}
{Using non-LTE modeling with RADEX, we derive the gas kinetic temperature
and spatial density, using as constraints the measured
para-H$_2$CO 3$_{21}$-2$_{20}$/3$_{03}$-2$_{02}$ and
para-H$_2$CO 3$_{03}$-2$_{02}$/C$^{18}$O 2-1 ratios.
Excluding the quiescent cloud N159S, where only one
para-H$_2$CO line could be detected, the gas kinetic
temperatures derived from the preferred
para-H$_2$CO 3$_{21}$-2$_{20}$/3$_{03}$-2$_{02}$ line ratios
range from 35 to 63 K with
an average of 47 $\pm$ 5 K (errors are unweighted standard deviations of the mean).
Spatial densities of the gas derived
from the para-H$_2$CO 3$_{03}$-2$_{02}$/C$^{18}$O 2-1 line ratios
yield 0.4 -- 2.9 $\times$ 10$^5$ cm$^{-3}$ with
an average of 1.5 $\pm$ 0.4 $\times$ 10$^5$ cm$^{-3}$.
Temperatures derived from the para-H$_2$CO line ratio are similar to those
obtained with the same method from Galactic star forming
regions and agree with results derived from CO in the dense regions
($n$(H$_2$) $>$ 10$^3$ cm$^{-3}$) of the LMC. A comparison of kinetic
temperatures derived from para-H$_2$CO with those from the dust also
shows good agreement. This suggests that the dust and para-H$_2$CO
are well mixed in the studied star forming regions. A comparison of
kinetic temperatures derived from
para-H$_2$CO 3$_{21}$-2$_{20}$/3$_{03}$-2$_{02}$ and NH$_3$(2,2)/(1,1)
shows, however, a drastic difference. In the star forming region N159W,
the gas temperature derived from the NH$_3$(2,2)/(1,1) line ratio is
$\sim$16 K \citep{Ott2010}, which is only half the temperature derived
from para-H$_2$CO and the dust. Furthermore, ammonia shows a very low
abundance in a 30$''$ beam. Apparently, ammonia only survives in the
most shielded pockets of dense gas not yet irradiated by UV photons,
while formaldehyde, less affected by photodissociation, is more
widespread and is also sampling regions more exposed to the radiation
of young massive stars. A correlation between the gas kinetic
temperatures derived from para-H$_2$CO and infrared luminosity,
represented by the 250 $\mu$m flux, suggests that the kinetic
temperatures traced by para-H$_2$CO are correlated with the ongoing
massive star formation in the LMC.}
{}
\maketitle

\section{Introduction}
The Large Magellanic Cloud (LMC), at a distance of $\sim$50 kpc
\citep{Pietrzynski2013}, is the nearest low metallicity
\citep{Rolleston2002} star-forming galaxy to the Milky Way.
The relatively face-on view offered by the LMC provides an ideal
perspective to study star formation, particularly massive star
formation associated with its numerous stellar clusters. In the LMC,
the FUV field is stronger than in the Milky Way \citep{Westerlund1990}.
It is thus well suited to study the properties of the
interstellar medium, the evolution of molecular clouds, and star
formation in an active low metallicity galaxy, also providing a
link to galaxies at high redshift.

The physical properties of the molecular gas in the LMC, in particular
the kinetic temperature, are not well constrained. The easily
thermalized and optically thick rotational CO transitions are good
temperature tracers of the cold and dense gas in local clouds.
Generally, however, they often suffer from a lack of information on the
beam filling factor in extragalactic clouds.
Multi-level observations of suitable molecules deliver the
most reliable temperature determinations.
The metastable lines of ammonia (NH$_3$) are frequently
used as the standard molecular cloud thermometer in
molecular clouds within our Galaxy and also in external galaxies
\citep{Ho1983,Walmsley1983,Danby1988,Henkel2000,Henkel2008,Weiss2001,
Mauersberger2003,Ao2011,Lebron2011,Ott2011,Wienen2012,Mangum2013a}.
However, the ammonia abundance can vary strongly in different
molecular environments (e.g., 10$^{-5}$ in dense,
molecular $``$hot cores$"$ around newly formed massive stars,
\citealt{Mauersberger1987}; 10$^{-8}$ in dark clouds,
\citealt{Benson1983,Chira2013}; 10$^{-10}$ in a massive star
forming cloud of the LMC, \citealt{Ott2010}) and is extremely
affected by a high UV flux. Thus, in a low metallicity environment
with high UV flux and a lack of shielding dust grains, ammonia
is of limited use as a reliable probe to trace the gas kinetic temperature.
To make this problem even more severe, the LMC is also a galaxy
with particulary low nitrogen abundance ($\sim$10\% solar,
\citealt{Wang2009,Ott2010}).

Formaldehyde (H$_2$CO) is a ubiquitous molecule in the
Galactic interstellar medium (ISM) of our and external galaxies
\citep{Downes1980,Cohen1981,Bieging1982,Cohen1983,Baan1986,Baan1990,
Baan1993,Henkel1991,Zylka1992,Huettemeister1997,Heikkila1999,
Wang2004,Wang2009,Mangum2008,Zhang2012,Mangum2013b,Ao2013,
Tang2013,Ginsburg2015,Ginsburg2016,Guo2016}.
H$_2$CO is thought to be formed on the surface
of dust grains by successive hydrogenation of CO
\citep{Watanabe2002,Woon2002,Hidaka2004}: CO $\rightarrow$ HCO $\rightarrow$ H$_2$CO.
Variations of the fractional abundance of H$_2$CO do not
exceed one order of magnitude. For example, the fractional
abundance of H$_2$CO is similar across
various sub regions of the well-studied Orion-KL nebula,
i.e., the $``$hot core$"$ and the $``$compact ridge$"$
\citep{Mangum1990,Mangum1993b,Caselli1993,Johnstone2003}.

Para-H$_2$CO has a rich variety of millimeter and submillimeter
transitions. Line ratios of para-H$_2$CO involving different
$K_{\rm a}$ ladders are good tracers of the kinetic
temperature, such as
para-H$_2$CO $J$$_{\rm KaKc}$ = 3$_{22}$-2$_{21}$/3$_{03}$-2$_{02}$,
4$_{23}$-3$_{22}$/4$_{04}$-3$_{03}$,
and 5$_{23}$-4$_{22}$/5$_{05}$-4$_{04}$, since the relative
populations of the $K_{\rm a}$ ladders of para-H$_2$CO are
predominantly governed by collisions \citep{Mangum1993a,Muhle2007}.
Among these para-H$_2$CO lines, the above three transitions
with rest frequencies of 218.222, 218.475, and 218.760 GHz,
respectively, are particularly useful for use a thermometer, because they are strong
enough for extragalactic observations and because they can be
measured simultaneously within a bandwidth of 1 GHz.
Temperature determined from these ratios are free from uncertainties related to pointing accuracy,
calibration or different beam sizes. Since the line emission is
optically thin and the levels are located up to 68 K above the
ground state, the line ratios are sensitive to gas kinetic
temperatures up to 50 K with relatively small uncertainties
\citep{Mangum1993a,Ao2013}. Para-H$_2$CO 3--2 line ratios have been
used to measure the molecular gas kinetic temperatures in our
Galactic center \citep{Qin2008,Ao2013,Johnston2014,Ginsburg2016,Immer2016},
star formation regions \citep{Mangum1993a,Mitchell2001,Watanabe2008,Tang2016},
as well as in external galaxies (e.g., \citealt{Muhle2007}).

Multitransition observations of molecular clouds in the LMC
\citep{Johansson1998,Heikkila1999,Israel2003,Kim2004,Bolatto2005,
Pineda2008,Mizuno2010,Minamidani2008,Minamidani2011,Fujii2014,Paron2016}
suggest that the molecular gas traced by CO may be warmer
and/or denser than in our Galaxy.
NH$_3$(1,1) and (2,2) lines have been surveyed toward seven
star-forming regions in the LMC by \cite{Ott2010} using the
Australia Telescope Compact Array (ATCA). Emission is only
detected in the massive star-forming region N159W. This
represents so far the only detection of NH$_3$ in the Magellanic Clouds.
The gas kinetic temperature derived from NH$_3$ (2,2)/(1,1)
is cold ($\sim$16 K), which is two times lower than the derived
dust temperature 30 -- 40 K
\citep{Heikkila1999,Bolatto2000,Gordon2014}. \cite{Ott2010}
also found a low fractional NH$_3$ abundance of
$\sim$4 $\times$ 10$^{-10}$, which is lower by 1.5 -- 5 orders
of magnitude than those observed in Galactic star-forming regions.
Previous observations of formaldehyde have been made in the LMC
(e.g., \citealt{Whiteoak1976,Johansson1994,Heikkila1999,Wang2009})
and H$_2$CO has been detected in many dense clumps. These
observations show that the fractional abundance of para-H$_2$CO
ranges from 1 to 6 $\times$ 10$^{-10}$, which agrees with the
values found in our Galactic molecular clouds
(e.g., \citealt{Gusten1983,Zylka1992,Johnstone2003,Ao2013,Tang2016}).

\begin{table}[t]
\tiny
\caption{Source coordinates, integration times, and epochs.}
\label{table:source}
\centering
\begin{tabular}
{ccccc}
\hline\hline 
Sources & RA(J2000) & DEC(J2000) & Int. Time & Obs. Date \\
& $^{h}$ {} $^{m}$ {} $^{s}$ & \degr {} \arcmin {} \arcsec & min &\\
\hline 
N159W &05:39:35.2  &-69:45:37.0 &229 & July-2008       \\
N113  &05:13:17.2  &-69:22:23.0 &91  & September-2008  \\
N44BC &05:22:02.8  &-67:57:45.8 &68  & May-2014        \\
30 Dor &05:38:49.3  &-69:04:44.0 &64  & May-2014        \\
N159S &05:40:02.8  &-69:50:32.7 &166 & May-July-2014   \\
N159E &05:40:04.4  &-69:44:34.0 &80  & July-2014       \\
\hline 
\end{tabular}
\end{table}

For this paper, we have carried out deep observations of the
six star forming regions 30 Dor, N44BC, N113, N159E, N159S,
and N159W in the LMC. Targeting three transitions of
para-H$_2$CO ($J$$_{K_AK_C}$ = 3$_{03}$-2$_{02}$, 3$_{22}$-2$_{21}$,
and 3$_{21}$-2$_{20}$) as well as C$^{18}$O 2-1 we
simultaneously determine kinetic temperatures and spatial densities
at high precision.
In Sections 2 and 3, we introduce our observations of the
para-H$_2$CO triplet and the data reduction, and describe the
main results. These are then discussed in Section 4. Our main
conclusions are summarized in Section 5.

\begin{figure*}[t]
\vspace*{0.2mm}
\begin{center}
\includegraphics[width=4.3cm]{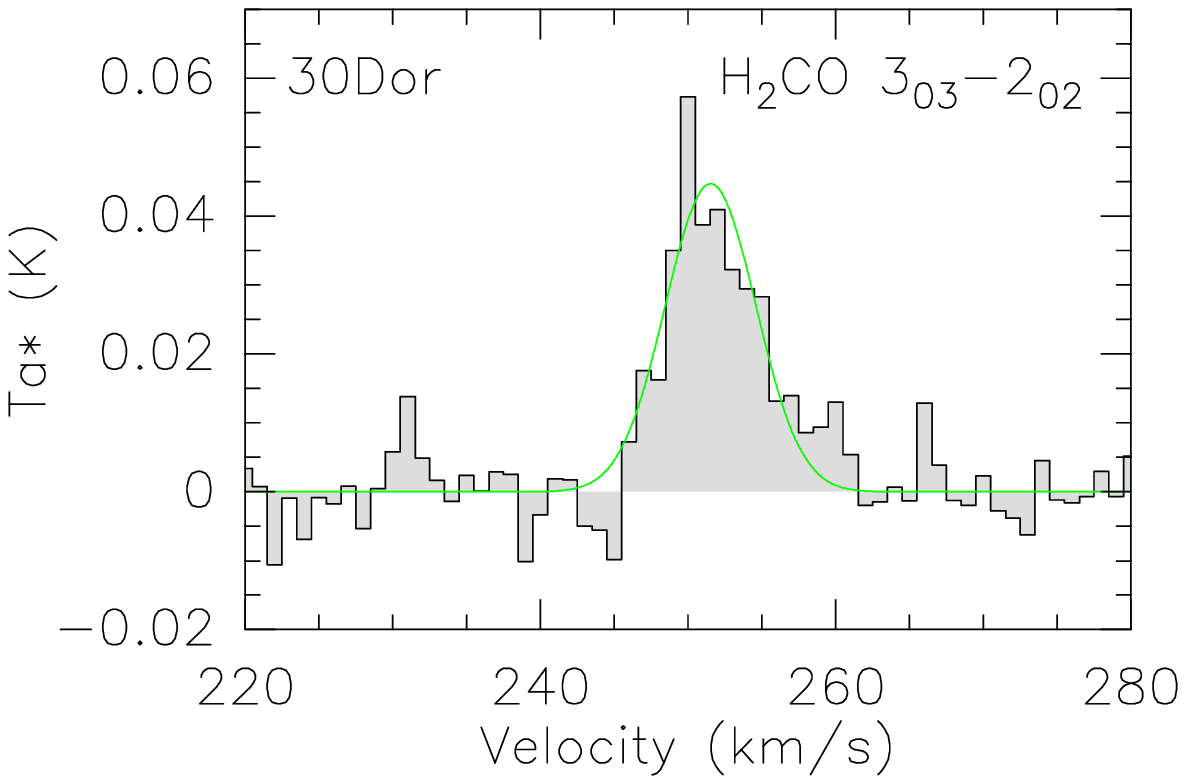}
\includegraphics[width=4.3cm]{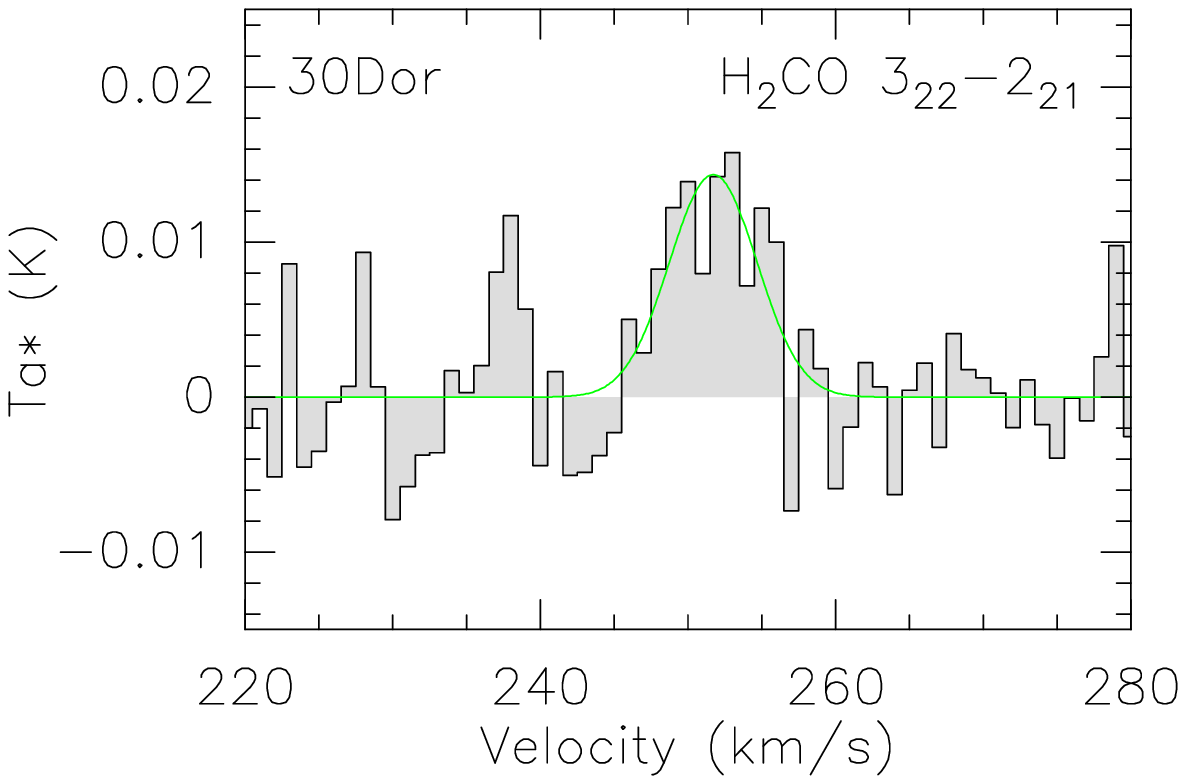}
\includegraphics[width=4.3cm]{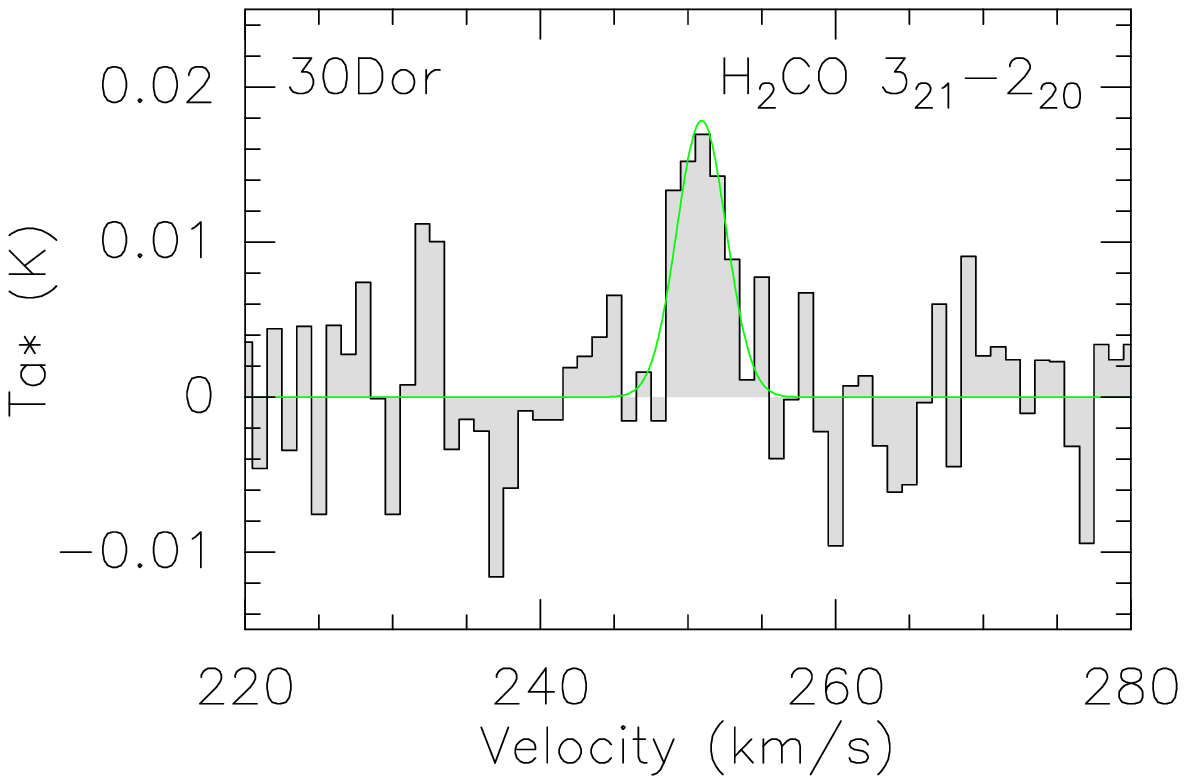}
\includegraphics[width=4.3cm]{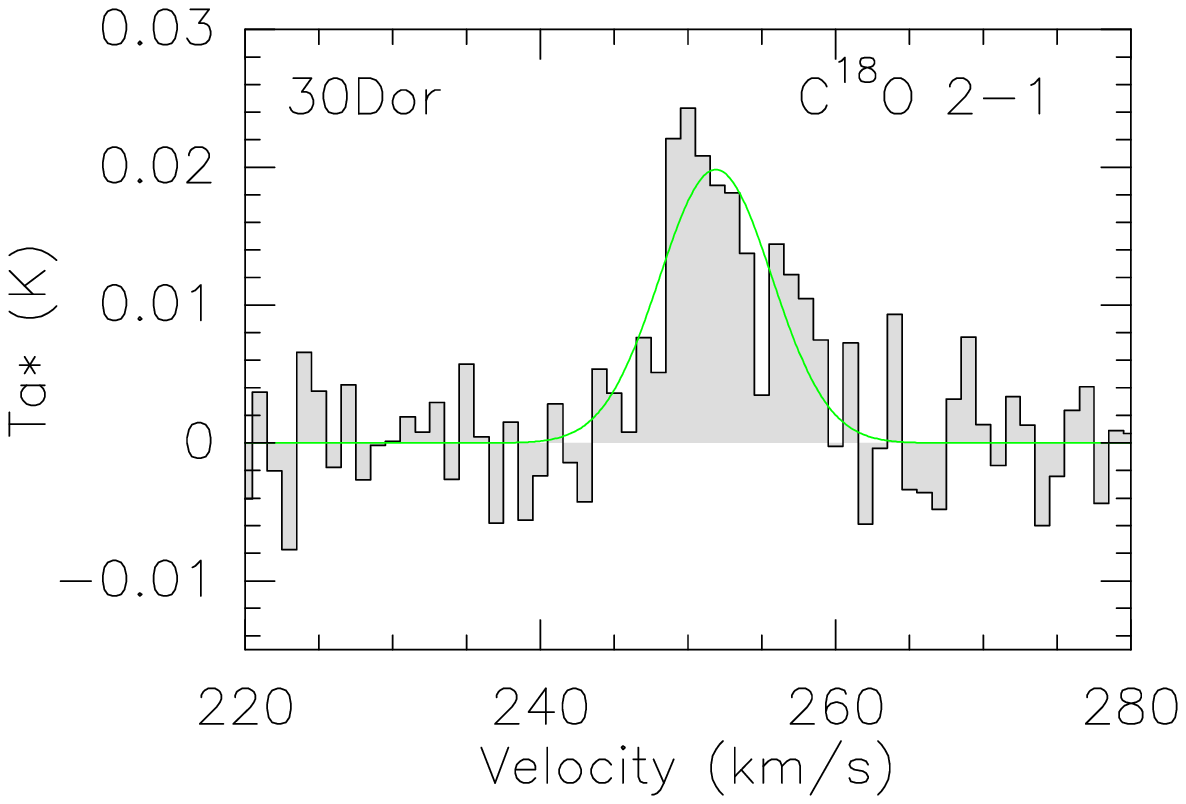}
\includegraphics[width=4.3cm]{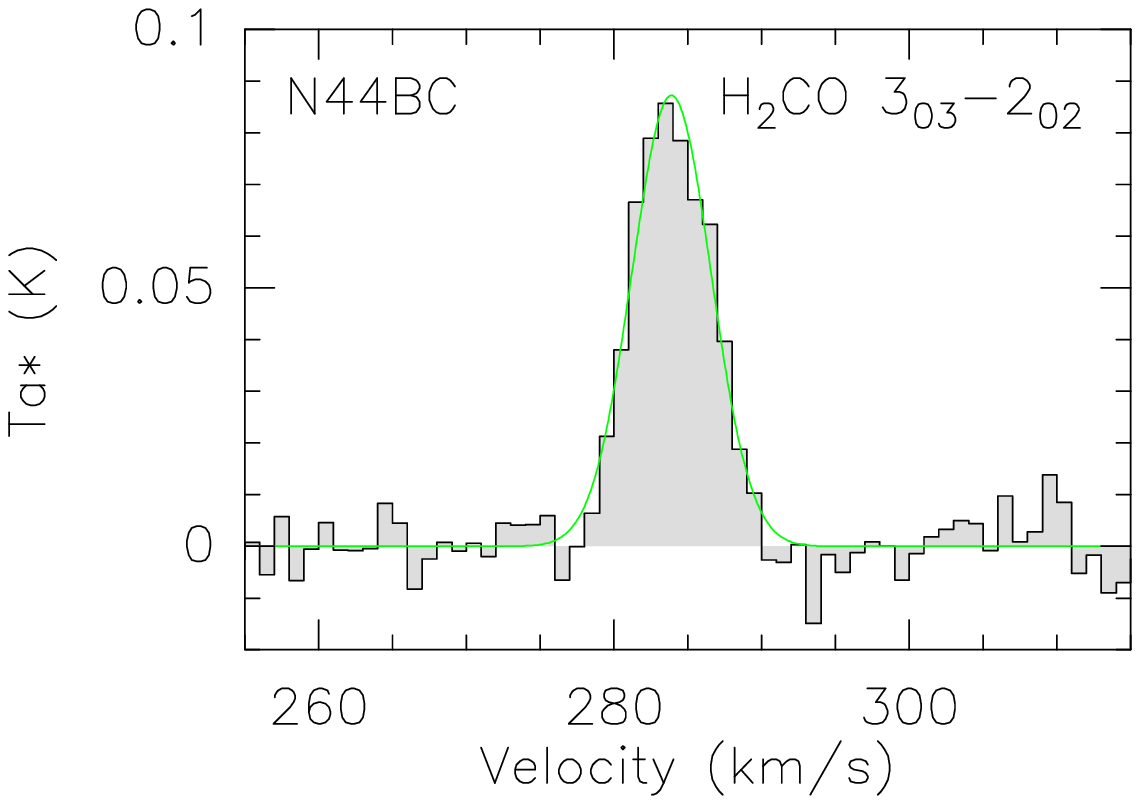}
\includegraphics[width=4.3cm]{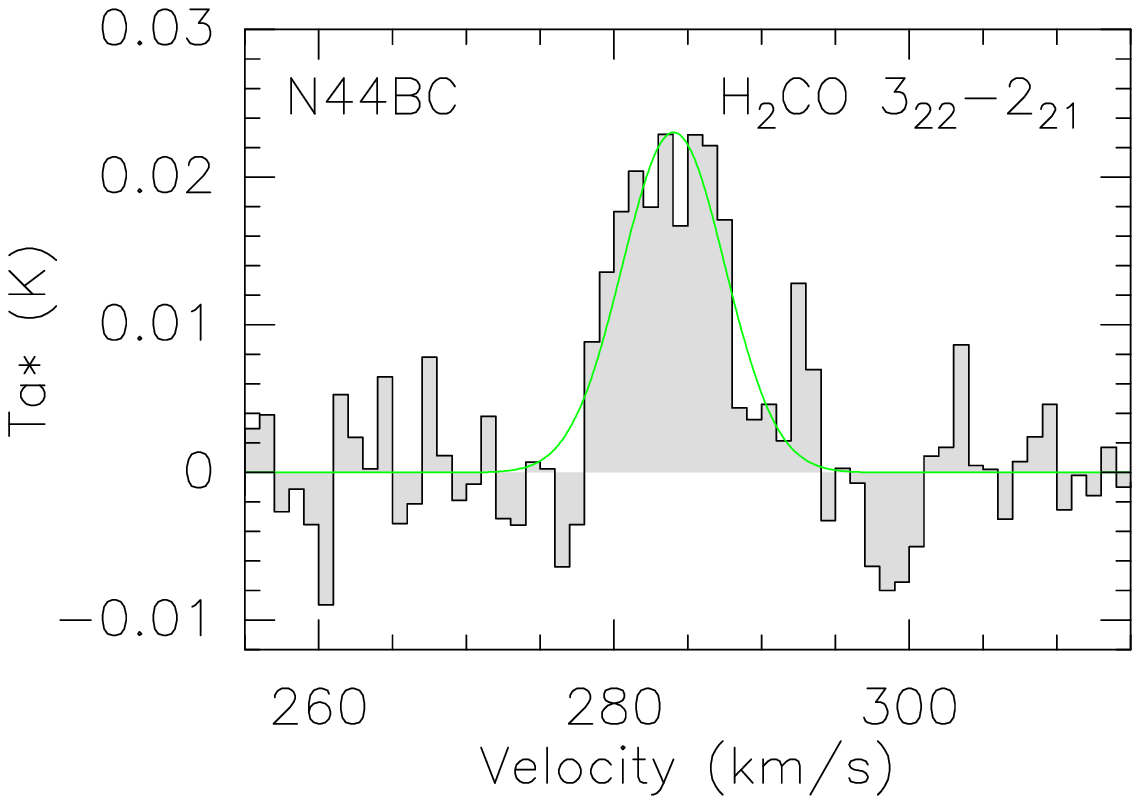}
\includegraphics[width=4.3cm]{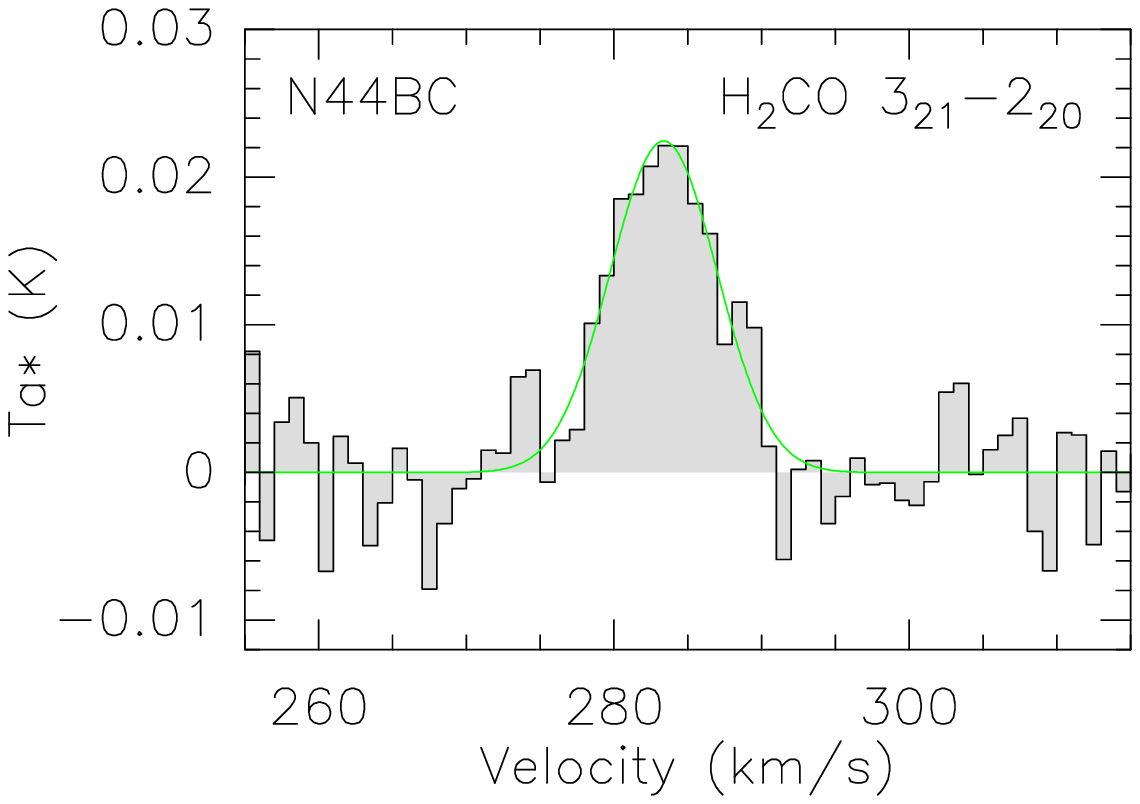}
\includegraphics[width=4.3cm]{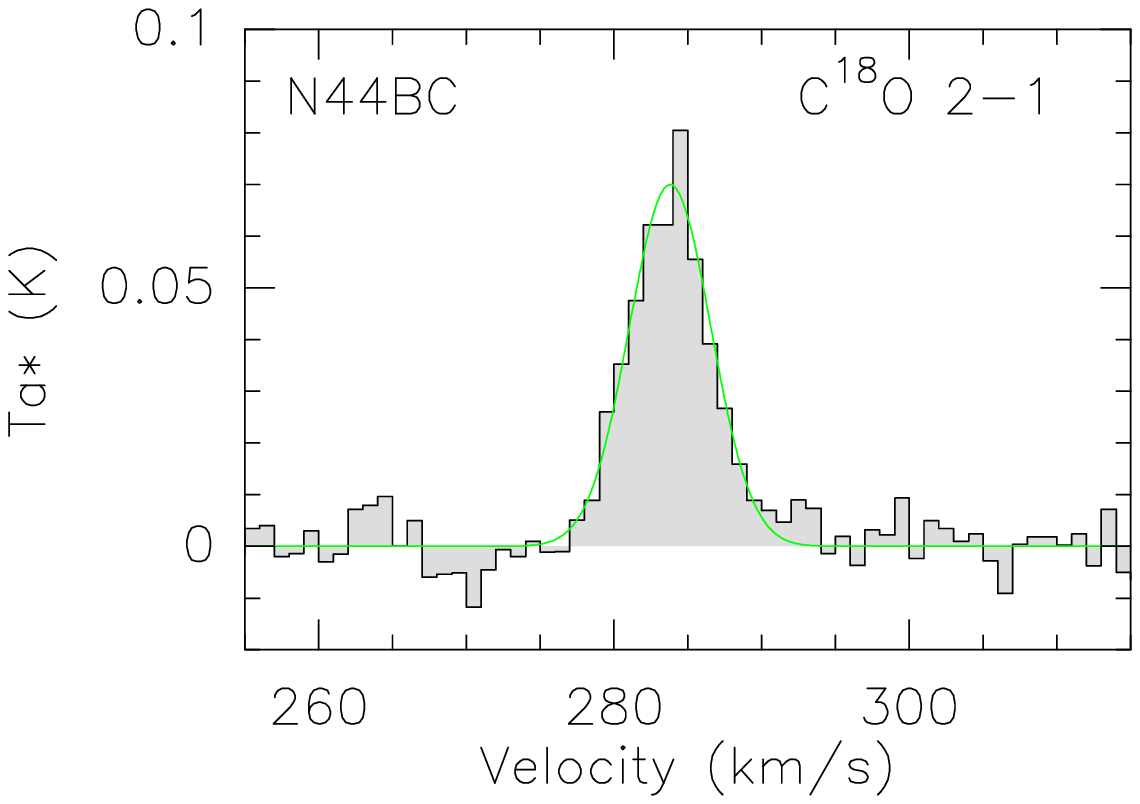}
\includegraphics[width=4.3cm]{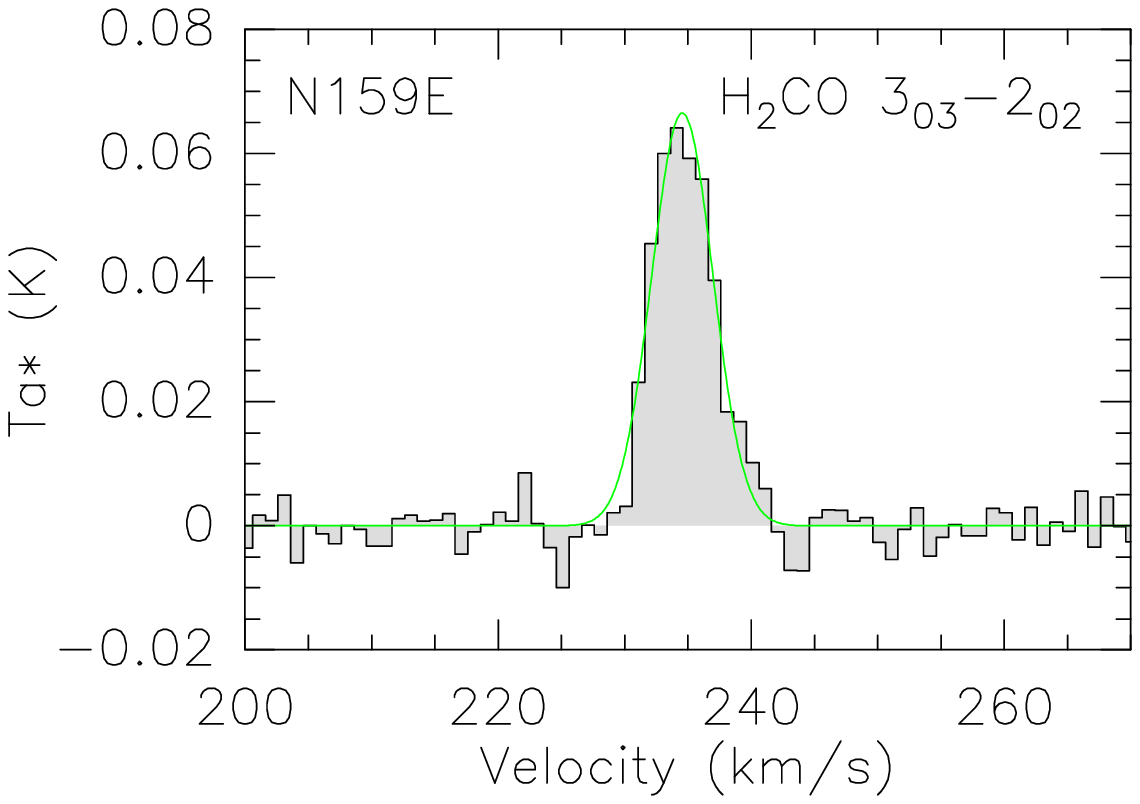}
\includegraphics[width=4.3cm]{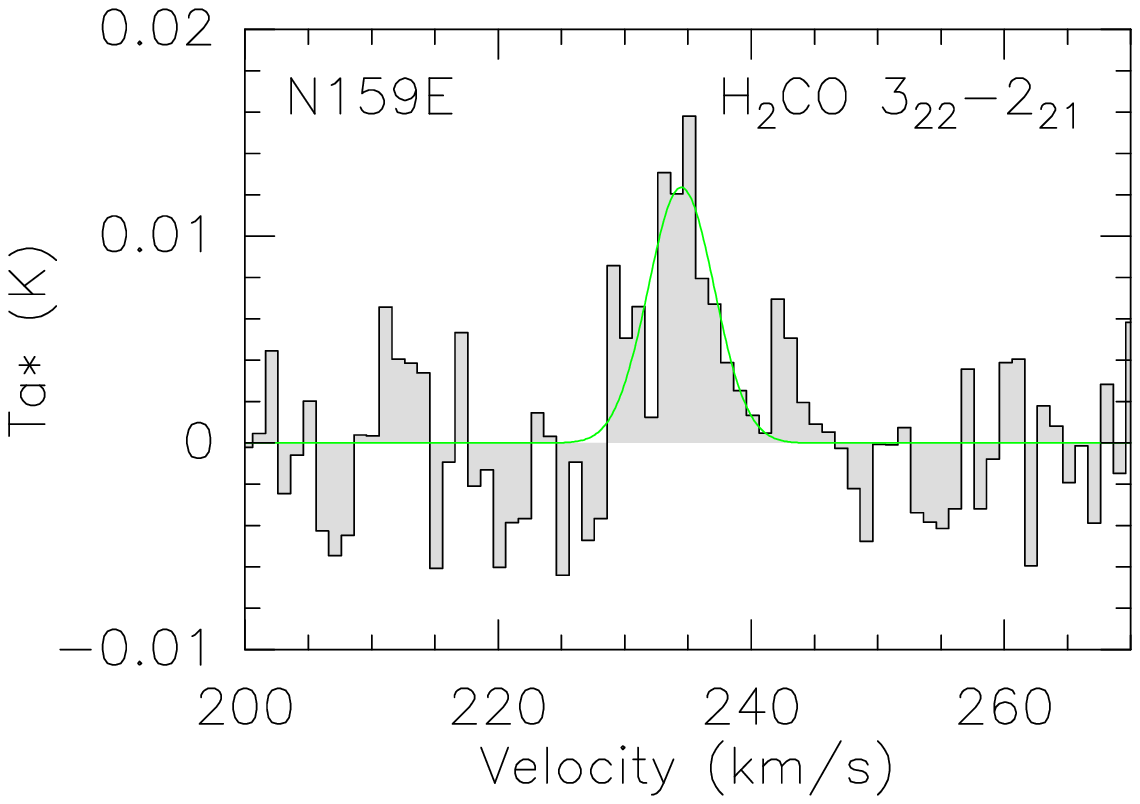}
\includegraphics[width=4.3cm]{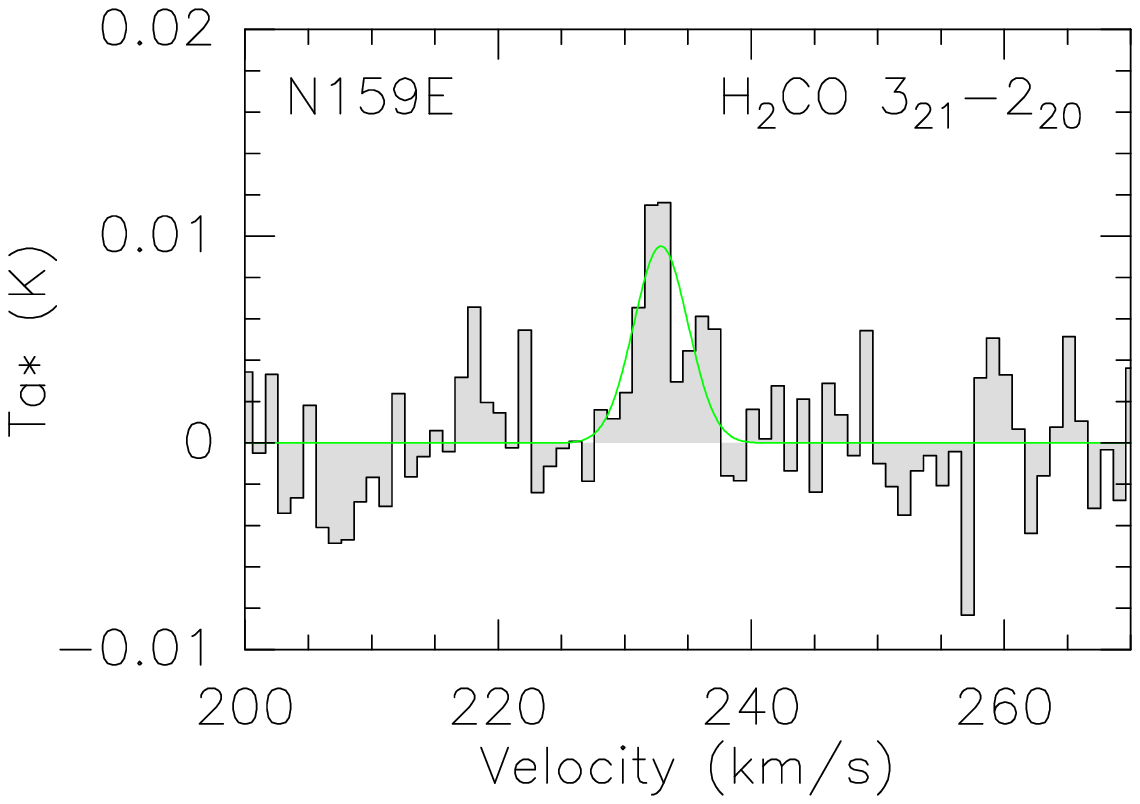}
\includegraphics[width=4.3cm]{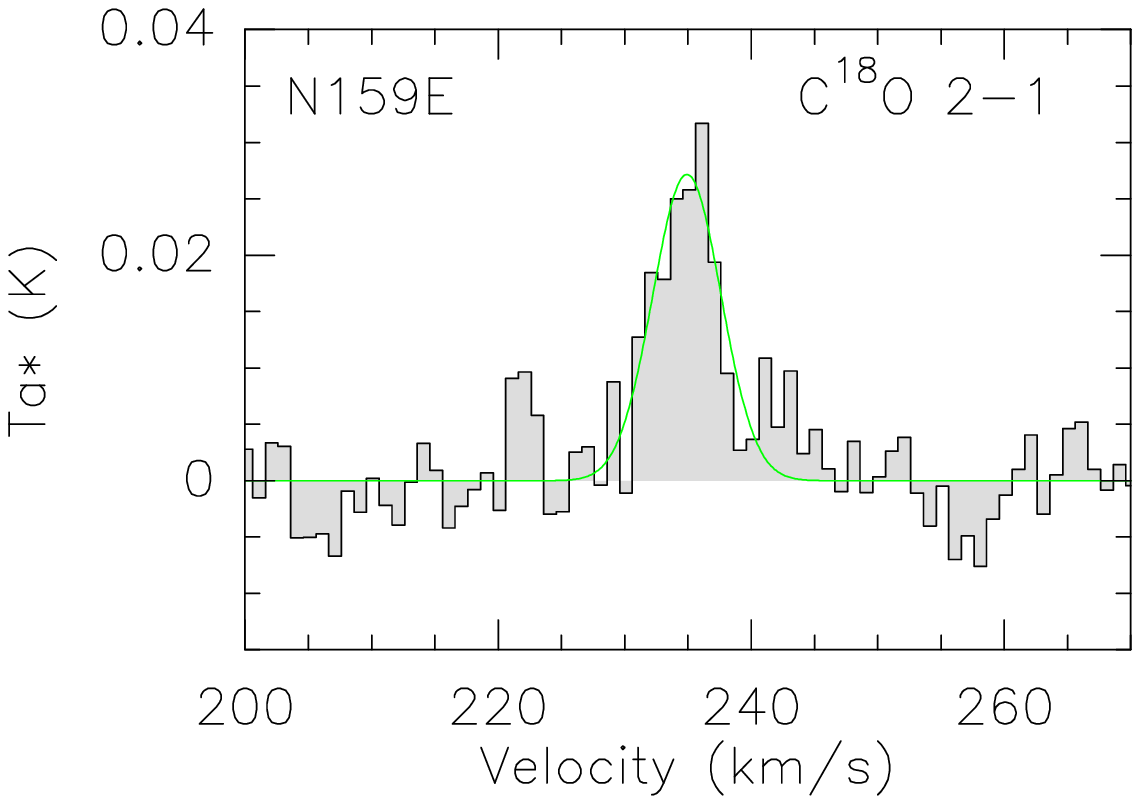}
\includegraphics[width=4.3cm]{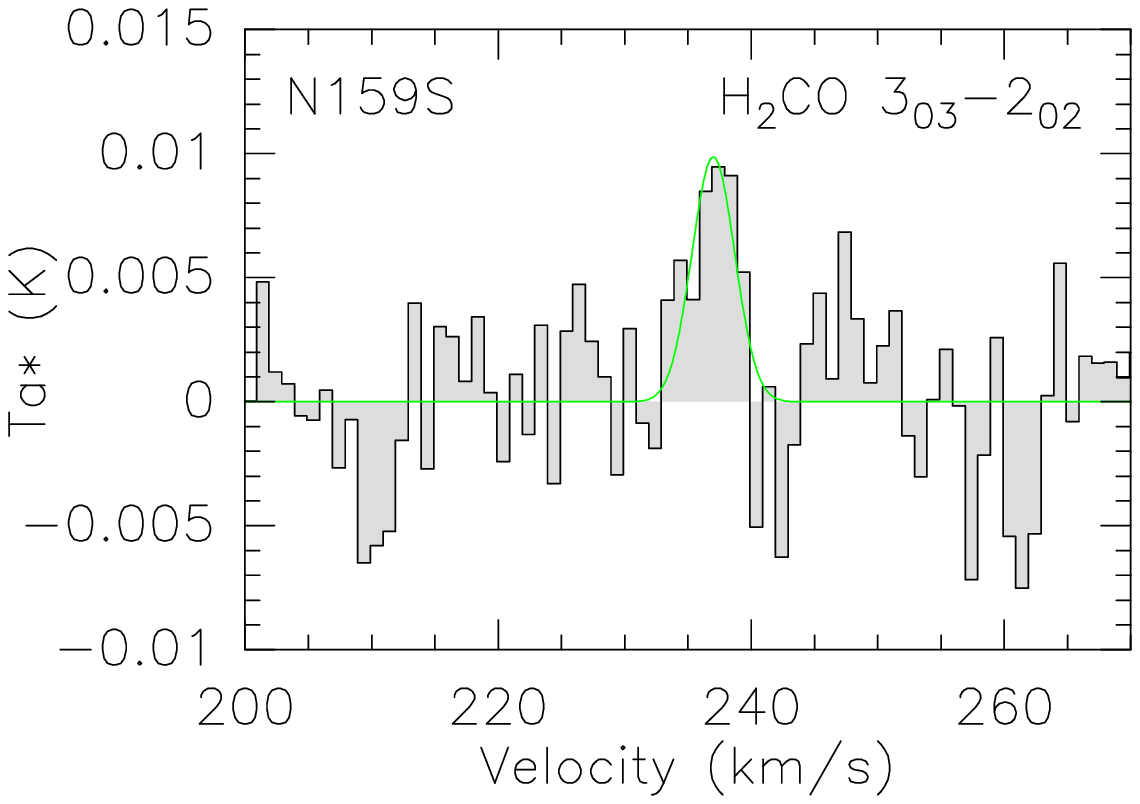}
\includegraphics[width=4.3cm]{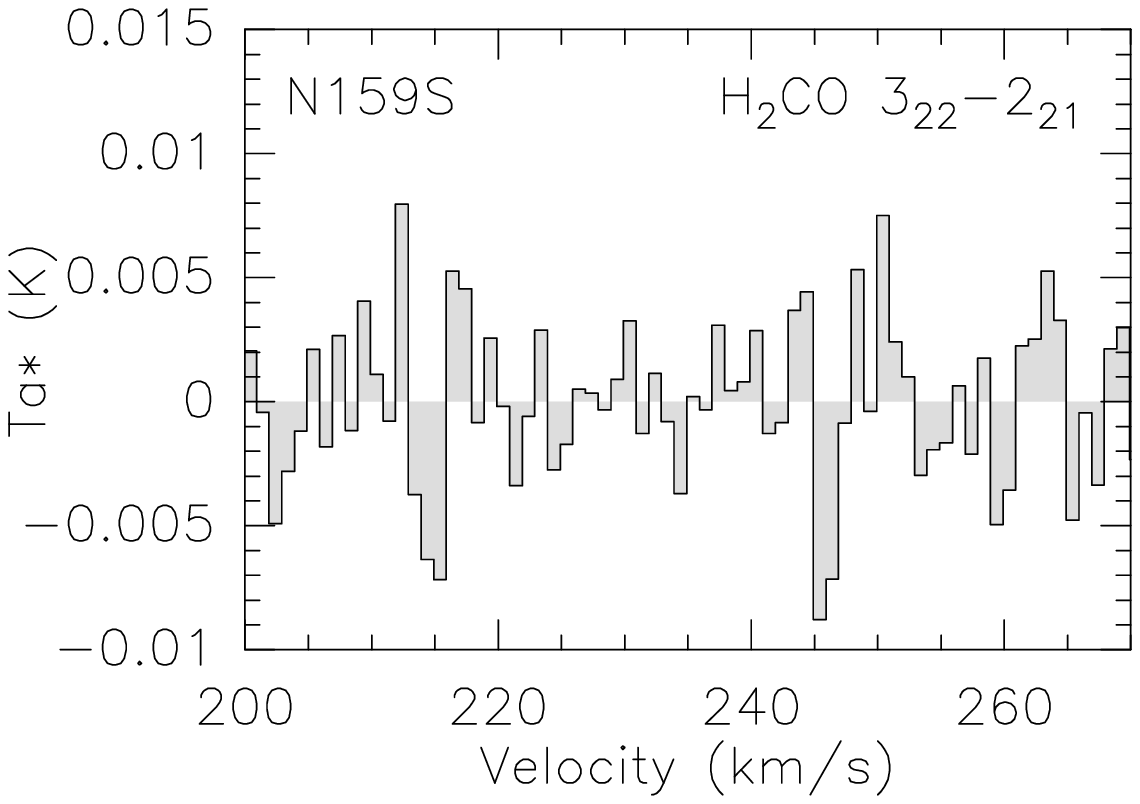}
\includegraphics[width=4.3cm]{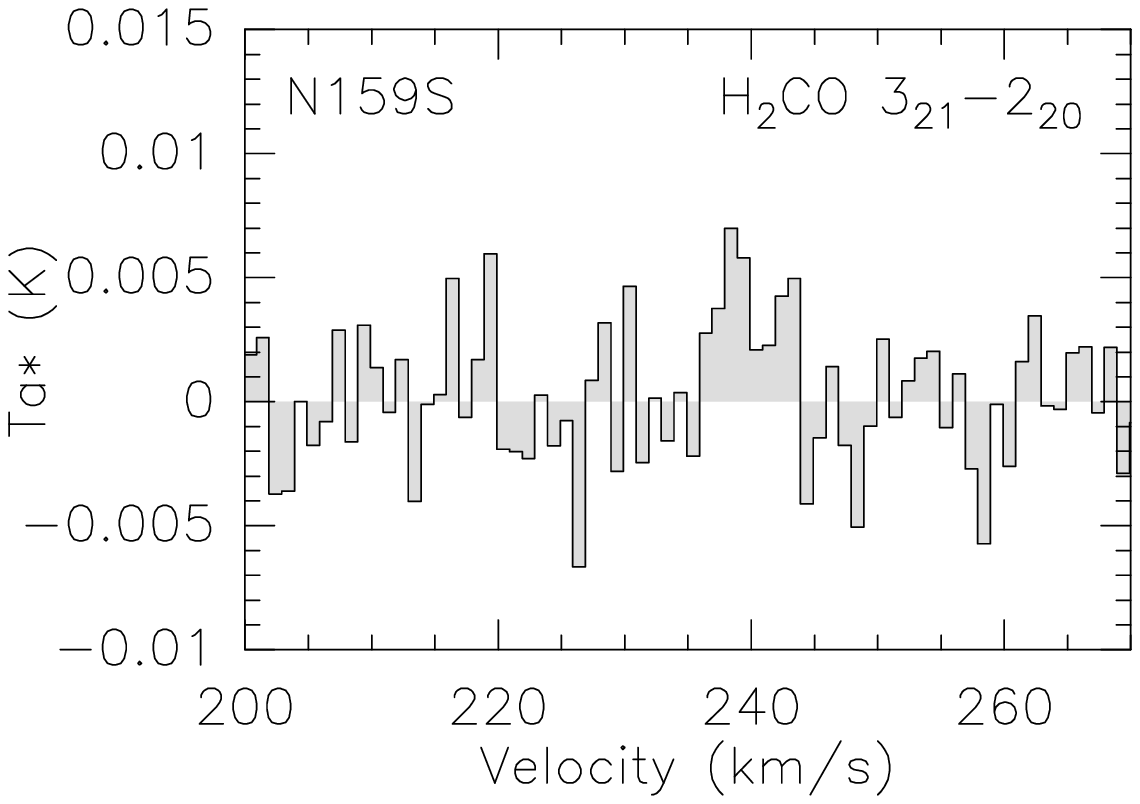}
\includegraphics[width=4.3cm]{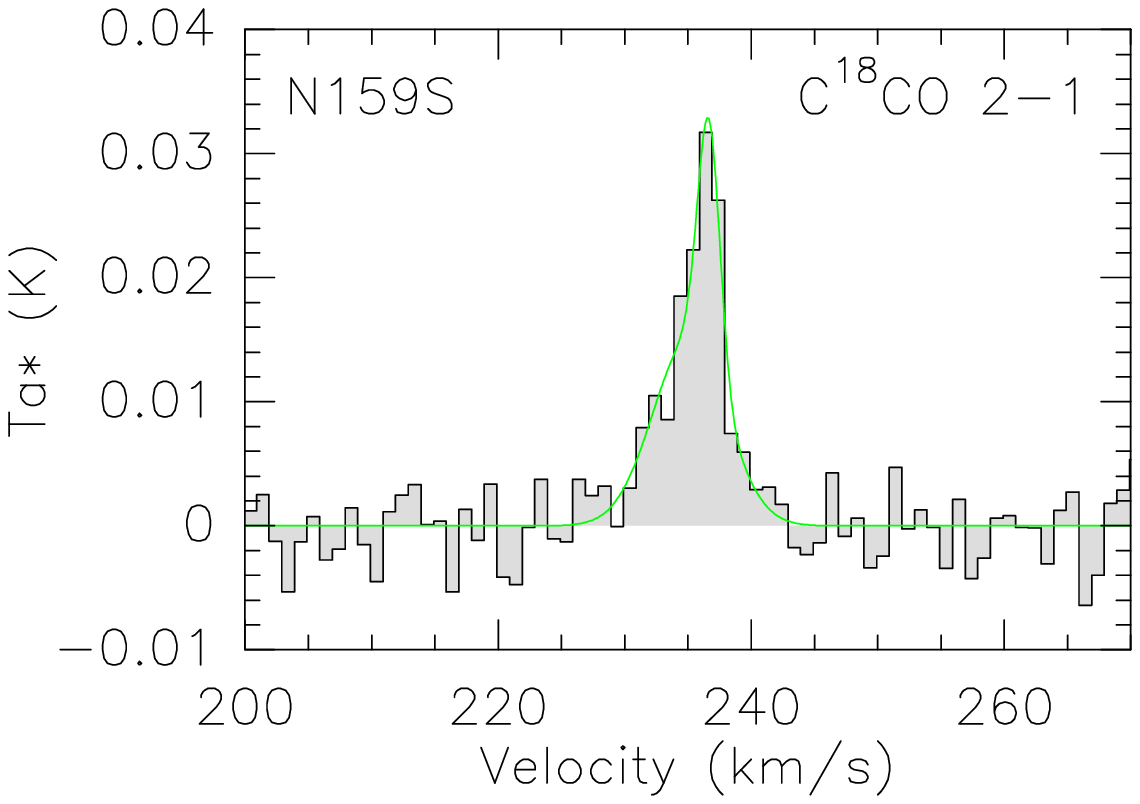}
\includegraphics[width=4.3cm]{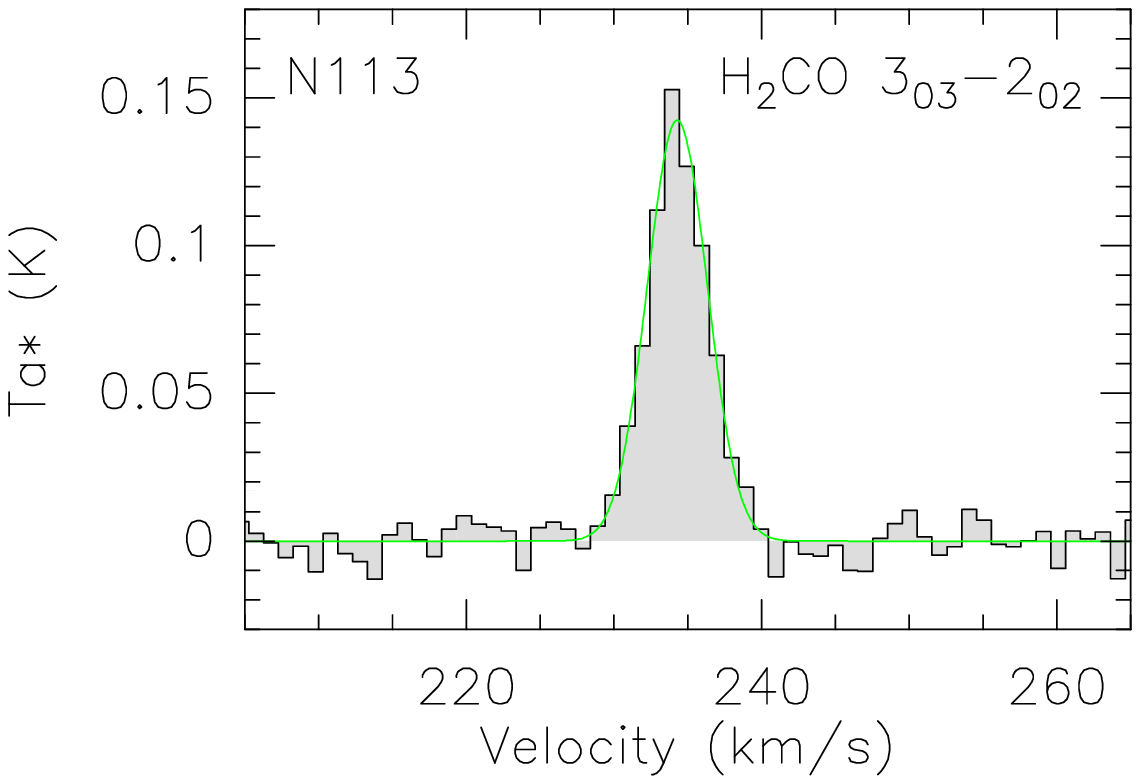}
\includegraphics[width=4.3cm]{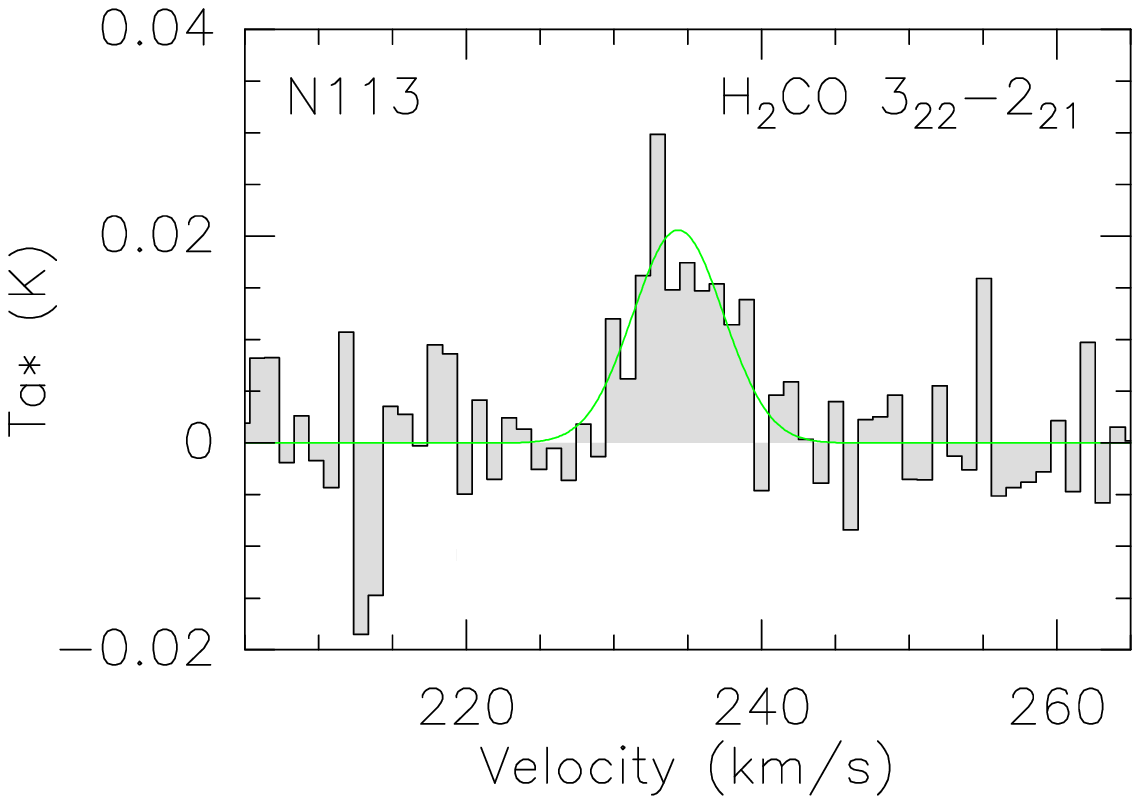}
\includegraphics[width=4.3cm]{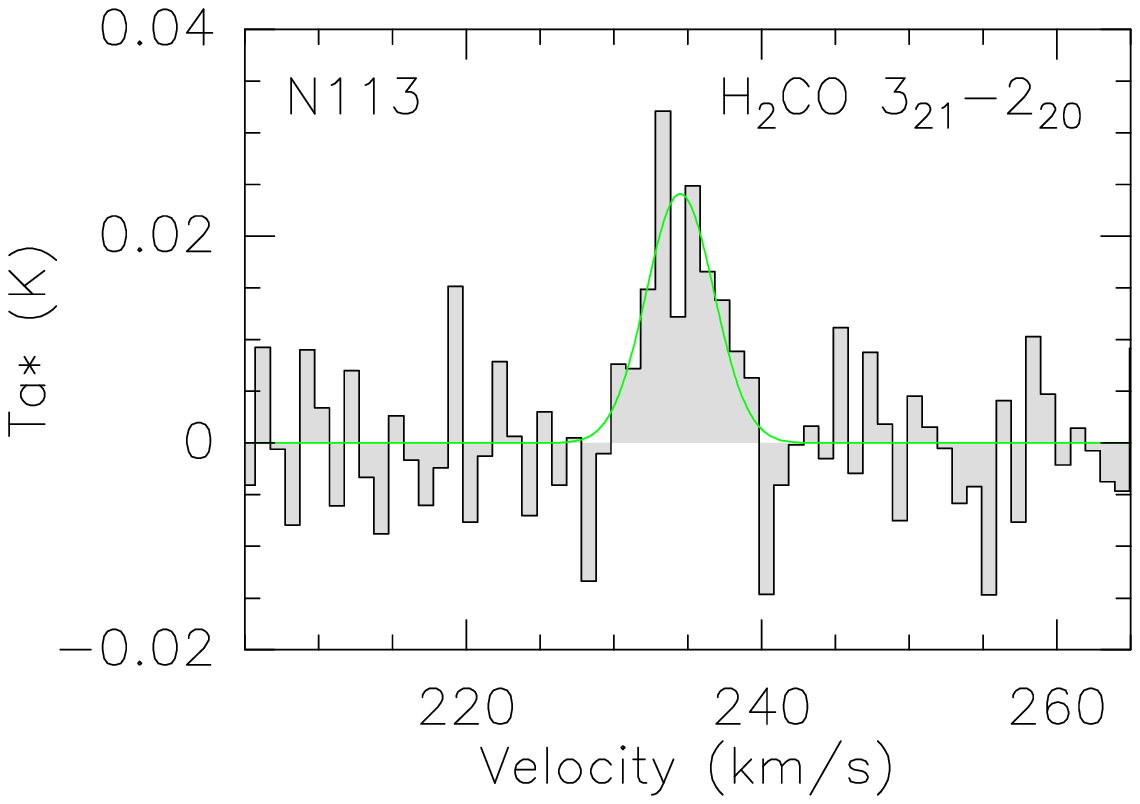}\\
\includegraphics[width=4.3cm]{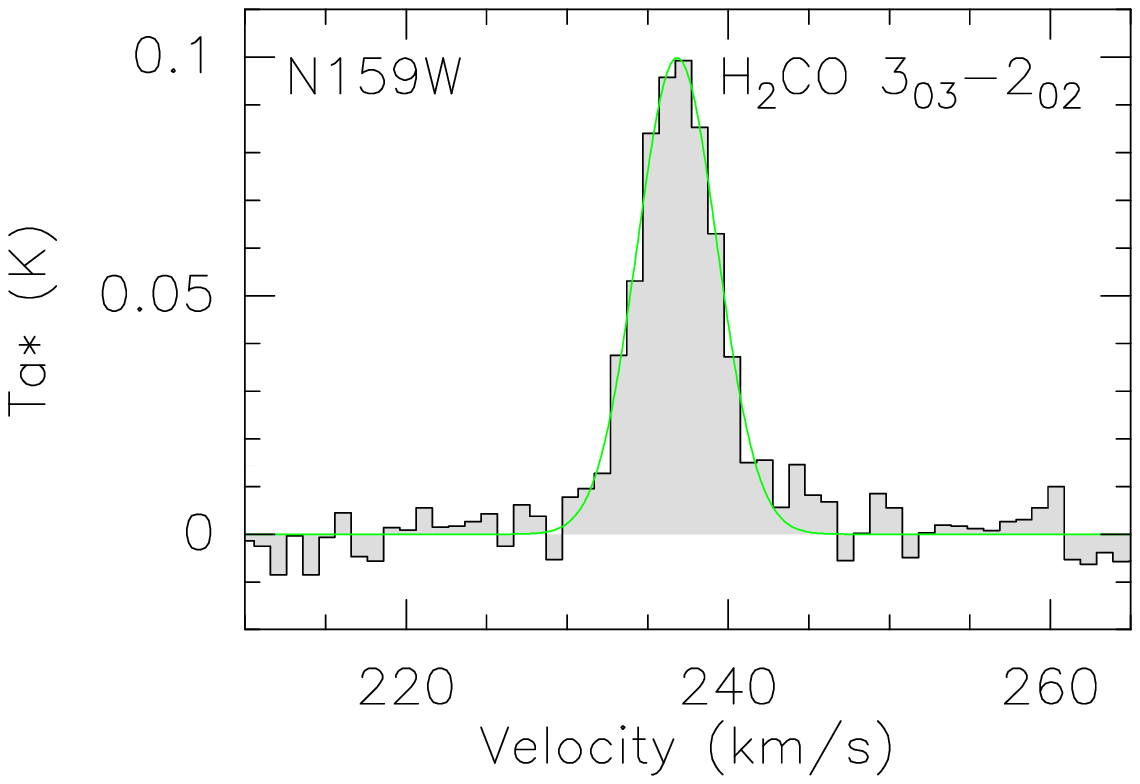}
\includegraphics[width=4.3cm]{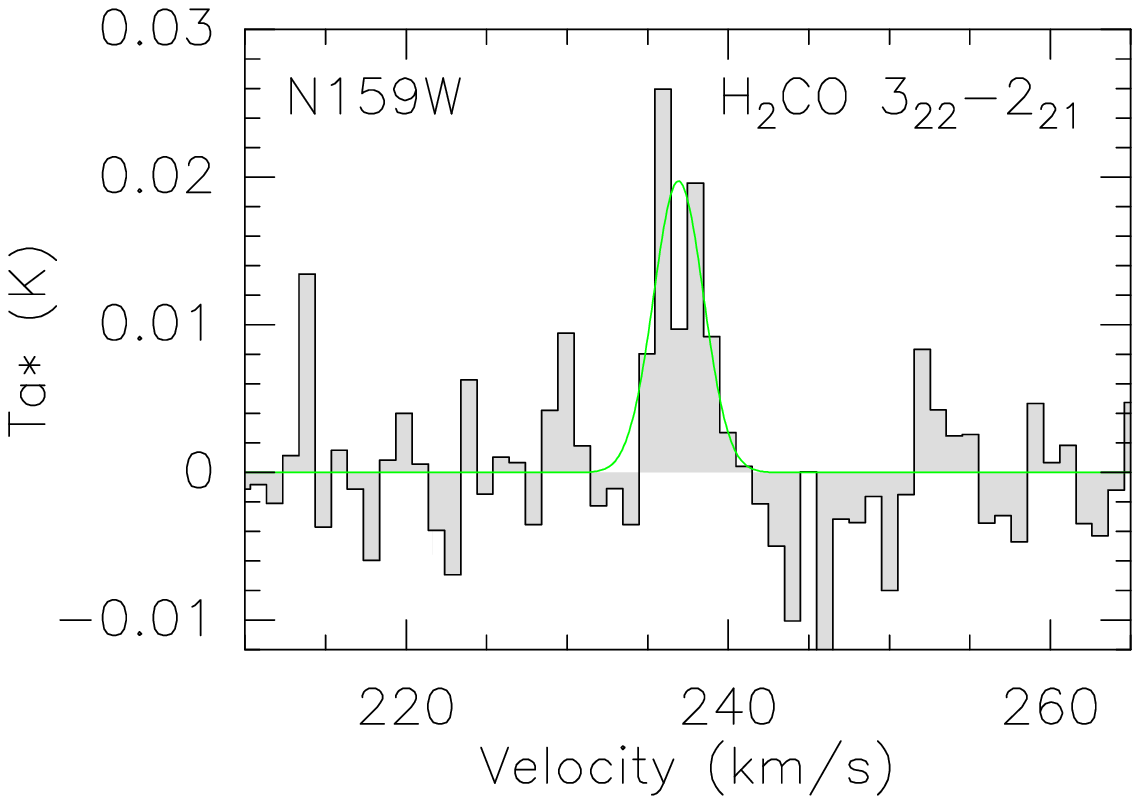}
\includegraphics[width=4.3cm]{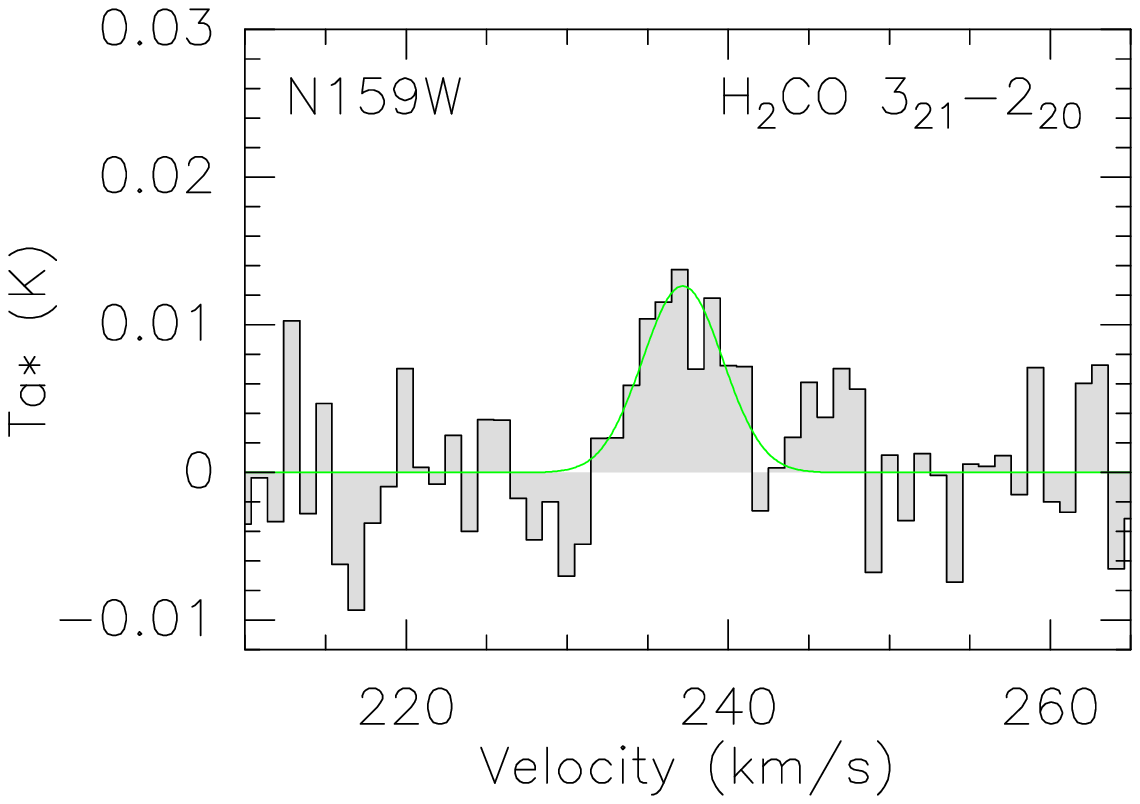}
\end{center}
\caption{Para-H$_2$CO (3$_{03}$-2$_{02}$, 3$_{22}$-2$_{21}$,
and 3$_{21}$-2$_{20}$) and C$^{18}$O (2-1) spectra.}
\label{fig:H2CO-C18O-spectra}
\end{figure*}

\section{Observations and data reduction}
Our observations were carried out in 2008 and 2014 (summarized
in Table \ref{table:source}) with the Atacama Pathfinder
EXperiment (APEX) 12 m telescope located on Chajnantor (Chile)
using the APEX-1 (SHeFI) receiver. The beam size is
$\sim$30$''$ ($\sim$7 pc at 50 kpc distance) at 218 GHz.
The main beam efficiency and the forward efficiency were
0.75 and 0.97, respectively. N113 and N159W were observed
in 2008 with an old Fast Fourier Transform Spectrometer (FFTS),
which consists of two units with a bandwidth of 1 GHz each and
a velocity resolution of 0.1675 km\,s$^{-1}$. The frequency is
centered at 218.480 GHz. Our data includes all three of the
218 GHz para-H$_2$CO lines. 30 Dor, N159E, N159S, and N44BC
were observed in 2014. Here we used the new eXtended bandwidth
Fast Fourier Transform Spectrometer (XFFTS) backend with two
spectral windows of 2.5 GHz bandwidth and a velocity resolution
of 0.1047 km\,s$^{-1}$. The central frequency is set at 218.550 GHz.
These data do not only include the three para-H$_2$CO lines
but also the 219.560 GHz C$^{18}$O (2-1) transition.

Toward each of the six star
forming regions in the LMC \cite{Wong2011} took a single pointing,
high sensitivity ($\sim$5 mK rms, $T_{\rm mb}$ scale,
beam size $\sim$30$''$) spectrum centered on its CO emission
peak which will be used to estimate $N$(H$_2$) in Section 4.1.
Ancillary C$^{18}$O 2-1 data have been published by \cite{Heikkila1998}
and \cite{Wang2009}. In addition, ammonia (NH$_3$) data from the LMC \citep{Ott2010} and
Herschel infrared data will also play an important role.

Data reduction for spectral lines was performed using
CLASS from the GILDAS package\footnote{\tiny http://www.iram.fr/IRAMFR/GILDAS}.
To enhance signal to noise
ratios (S/N) in individual channels, we smoothed contiguous
channels to a velocity resolution of $\sim$1.0 km\,s$^{-1}$.
Sources observed are listed in Table \ref{table:source}.

\begin{table*}[t]
\caption{Para-H$_2$CO and C$^{18}$O spectral parameters$^a$.}
\centering
\begin{tabular}
{cccccccc}
\hline\hline 
Sources & Molecule & Transition & $\int$$T$$_{\rm mb}$d$v$  & $V_{\rm lsr}$ & FWHM$^b$ & $T$$_{\rm mb}$ & rms\\
&  & & K km\,s$^{-1}$ & km\,s$^{-1}$ & km\,s$^{-1}$ & K &mK\\
\hline
30 Dor & H$_2$CO   &3$_{03}$ -- 2$_{02}$ & 0.401 (0.026) & 250.92 (0.21) & 6.12 (0.47) & 0.061 & 6.8\\
       & H$_2$CO   &3$_{22}$ -- 2$_{21}$ & 0.114 (0.023) & 251.61 (0.70) & 6.48 (1.19) & 0.017 & 6.5\\
       & H$_2$CO   &3$_{21}$ -- 2$_{20}$ & 0.109 (0.021) & 251.01 (0.41) & 4.29 (1.00) & 0.023 & 6.4\\
       & C$^{18}$O &2 -- 1               & 0.190 (0.026) & 251.25 (0.48) & 7.09 (1.34) & 0.025 & 5.6\\
\hline
N113   & H$_2$CO   &3$_{03}$ -- 2$_{02}$ & 0.930 (0.029) & 234.25 (0.07) & 4.86 (0.17) & 0.181 & 7.0\\
       & H$_2$CO   &3$_{22}$ -- 2$_{21}$ & 0.217 (0.035) & 234.40 (0.58) & 7.05 (1.20) & 0.029 & 6.9\\
       & H$_2$CO   &3$_{21}$ -- 2$_{20}$ & 0.213 (0.025) & 234.36 (0.33) & 5.69 (0.68) & 0.035 & 7.3\\
\hline
N44BC  & H$_2$CO   &3$_{03}$ -- 2$_{02}$ & 0.775 (0.026) & 283.86 (0.11) & 6.37 (0.23) & 0.114 & 6.5\\
       & H$_2$CO   &3$_{22}$ -- 2$_{21}$ & 0.155 (0.019) & 283.97 (0.45) & 6.98(0.88)  & 0.021 & 5.2\\
       & H$_2$CO   &3$_{21}$ -- 2$_{20}$ & 0.152 (0.021) & 283.45 (0.47) & 6.50 (0.90) & 0.022 & 5.1\\
       & C$^{18}   $O&2 -- 1             & 0.558 (0.022) & 283.78 (0.12) & 6.08 (0.28) & 0.086 & 6.2\\
\hline
N159W  & H$_2$CO   &3$_{03}$ -- 2$_{02}$ & 0.788 (0.019) & 236.78 (0.07) & 5.82 (0.17) & 0.127 & 6.2\\
       & H$_2$CO   &3$_{22}$ -- 2$_{21}$ & 0.101 (0.017) & 237.00 (0.30) & 3.65 (0.70) & 0.026 & 6.4\\
       & H$_2$CO   &3$_{21}$ -- 2$_{20}$ & 0.105 (0.017) & 236.98 (0.53) & 5.94 (1.00) & 0.017 & 5.2\\
\hline
N159E  & H$_2$CO   &3$_{03}$ -- 2$_{02}$ & 0.555 (0.017) & 234.53 (0.09) & 5.84 (0.21) & 0.090 & 4.9\\
       & H$_2$CO   &3$_{22}$ -- 2$_{21}$ & 0.112 (0.018) & 234.45 (0.47) & 6.46 (1.37) & 0.016 & 4.3\\
       & H$_2$CO   &3$_{21}$ -- 2$_{20}$ & 0.073 (0.016) & 232.94 (0.59) & 5.36 (1.53) & 0.013 & 3.9\\
       & C$^{18}$O &2 -- 1               & 0.231 (0.019) & 234.85 (0.23) & 6.17 (0.66) & 0.035 & 4.8\\
\hline
N159S  & H$_2$CO   &3$_{03}$ -- 2$_{02}$ & 0.056 (0.010) & 235.87 (0.45) & 4.45 (0.80) & 0.012 & 3.9\\
       & H$_2$CO   &3$_{22}$ -- 2$_{21}$ & $<$0.005      & ...           & ...         & ...   & 4.9\\
       & H$_2$CO   &3$_{21}$ -- 2$_{20}$ & $<$0.005      & ...           & ...         & ...   & 3.4\\
       & C$^{18}$O &2 -- 1               & 0.161 (0.006) & 236.35 (0.07) & 3.27 (0.19) & 0.047 & 3.4\\
       &           &                     & 0.047 (0.006) & 232.39 (0.30) & 3.40 (0.52) & 0.013 & 3.4\\
\hline
 \end{tabular}
\label{table:H2CO-C18O}
\tablefoot{\\
(a) Values in parenthesis are standard deviations from Gaussian fits using CLASS as part of the GILDAS software.\\
(b) Full width to half maximum line width.}
   \end{table*}

\section{Results}
The three para-H$_2$CO lines are detected in all sources
except N159S. There, only the strongest para-H$_2$CO line,
the 3$_{03}$-2$_{02}$ transition, is detected. C$^{18}$O (2-1),
measured in 30 Dor, N159E, N159S, and N44BC, is detected
in all sources. The para-H$_2$CO and C$^{18}$O line
spectra are presented in Figure \ref{fig:H2CO-C18O-spectra}.
Line parameters are listed in Table \ref{table:H2CO-C18O},
where velocity-integrated intensity, $\int$$T$$_{\rm mb}$d$v$, local
standard of rest velocity, $V_{\rm lsr}$, full width to half
maximum line width, FWHM, peak main beam brightness temperature,
$T$$_{\rm mb}$, and rms noise, were obtained from Gaussian fits.

\subsection{Kinetic temperature and spatial density}
To determine gas kinetic temperatures and spatial densities,
we use the RADEX non-LTE model \citep{van der Tak2007}
offline code\footnote{%
  \tiny
http://var.sron.nl/radex/radex.php} with H$_2$CO collision
rates from \cite{Wiesenfeld2013} and C$^{18}$O collision
rates from \cite{Yang2010}.
The RADEX code needs five input parameters: background
temperature, kinetic temperature, H$_2$ density, column density,
and line width. For the background temperature, we adopt 2.73 K.
Model grids for the para-H$_2$CO and C$^{18}$O lines encompass
30 densities ($n$(H$_2$) = 10$^3$ -- 10$^7$ cm$^{-3}$) and 30
temperatures ranging from 10 to 110 K.
For the line width, we use the observed line width value
(Table \ref{table:H2CO-C18O}).
The total beam averaged column density of C$^{18}$O can be obtained from the
$J$ = 2--1 integrated intensity following \cite{Batrla2003}.
\begin{equation}
N({\rm C^{18}O}) = 5.3 \times 10^{14} \int T_{\rm mb} ({\rm C^{18}O}, J=2-1){\rm d}v,
\end{equation}
where $\int T_{\rm mb} ({\rm C^{18}O}, J=2-1){\rm d}v$ is the
C$^{18}$O (2-1) integrated intensity.
The results are listed in Table \ref{table:Tkin}.
In the LMC, the gas is well mixed so that isotope ratios are
pretty much the same throughout the galaxy
\citep{Johansson1994,Chin1997,Heikkila1998,Heikkila1999,Wang2009}.
In the local ISM, the $^{16}$O/$^{18}$O ratio
is approximately 500 \citep{Wilson1994}. We have, however, to keep in
mind that fractional abundances in the LMC differ from those in the local ISM.
In the LMC, $^{18}$O is underabundant
with respect to $^{16}$O by about a factor of
$\sim$4 ($^{16}$O/$^{18}$O $\sim$2000 in the LMC, \citealt{Chin1999,Wang2009}),
while $^{18}$O is underabundant by a factor of $\sim$2.4 with
respect to $^{17}$O (locally, $^{18}$O/$^{17}$O $\sim$4.1,
\citealt{Zhang2007,Wouterloot2008}; in the LMC $\sim$1.7,
\citealt{Heikkila1998,Wang2009}). Therefore, $N$(H$_2$)/$N$(C$^{18}$O)
ratios should be higher by a factor of $\sim$3 in the LMC with
respect to the local ISM. This correlation will be used to derive
H$_2$ column densities. The results for the H$_2$ column density
are listed in Table \ref{table:Tkin}.
Assuming a CO to H$_2$ conversion factor of
$X_{CO}$ $\sim$ 4 $\times$ 10$^{20}$ cm$^{-2}$ (K km\,s$^{-1}$)$^{-1}$
for the LMC \citep{Pineda2009,Wang2009}, we can also derive a column
density of $N$(H$_2$) from the CO(1-0) integrated intensity
moment map reported by \cite{Wong2011}. The results of $N$(H$_2$)
are listed in Table \ref{table:Tkin}. The H$_2$ column densities
derived from C$^{18}$O and CO are consistent with in the estimated uncertainty
by a factor of two.

Previous observational results on
transitions of C$^{18}$O (1-0, 2-1) and para-H$_2$CO (2$_{02}$-1$_{01}$,
3$_{03}$-2$_{02}$, 3$_{22}$-2$_{21}$ and 3$_{21}$-2$_{20}$) using the 15 m
Swedish-ESO Submillimetre Telescope (SEST)
show that the column density ratio of $N$(C$^{18}$O)/$N$(para-H$_2$CO)
is $\sim$ 100 in dense clumps of the LMC \citep{Heikkila1998,Heikkila1999,Wang2009}.
Assuming that the $N$(C$^{18}$O)/$N$(para-H$_2$CO) ratio is the same in our sample,
we estimate the column density of para-H$_2$CO from the $N$(C$^{18}$O) column density.
The results are listed in Table \ref{table:Tkin}.

\begin{figure*}[t]
\vspace*{0.2mm}
\begin{center}
\includegraphics[width=6cm]{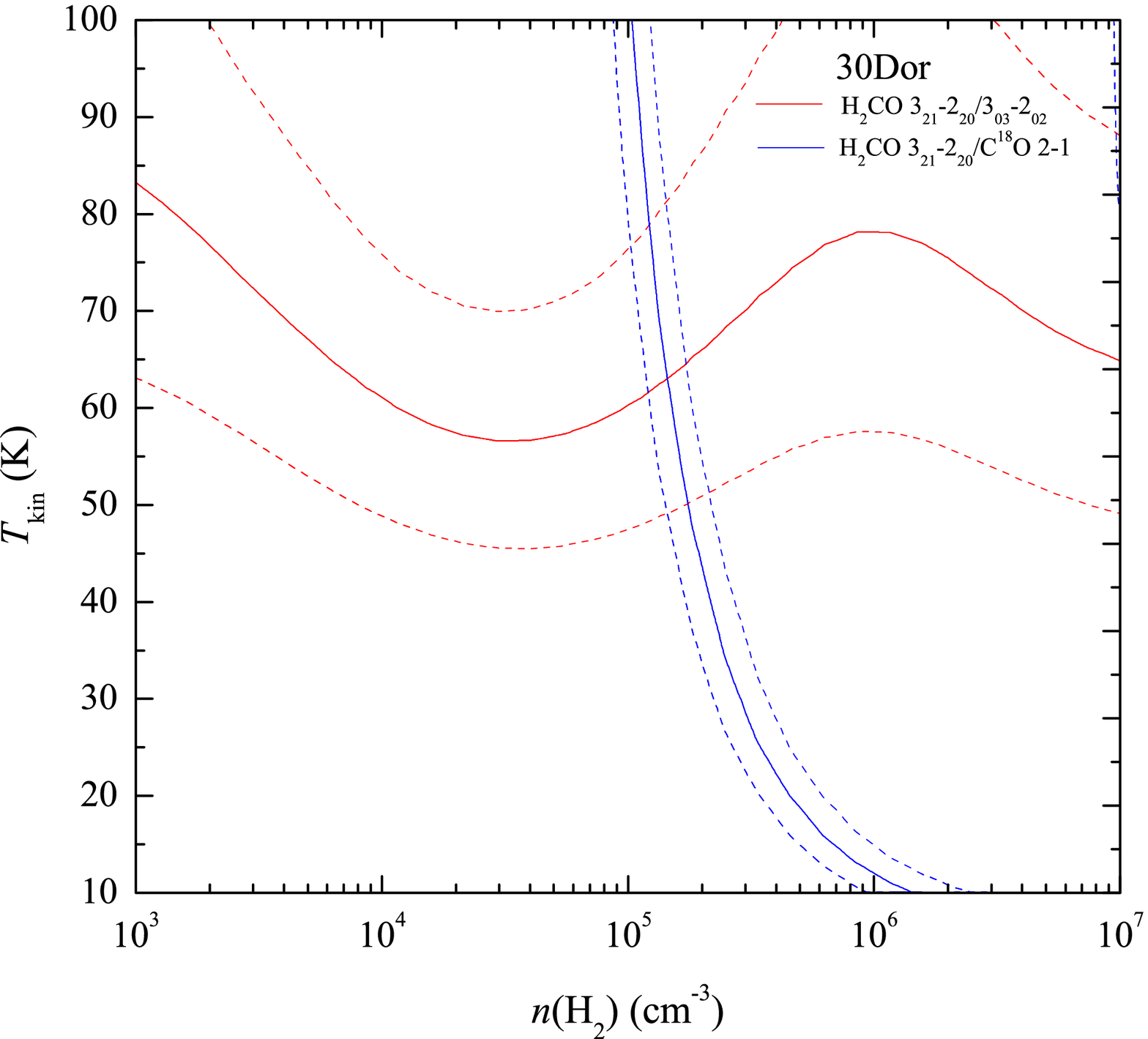}
\includegraphics[width=6cm]{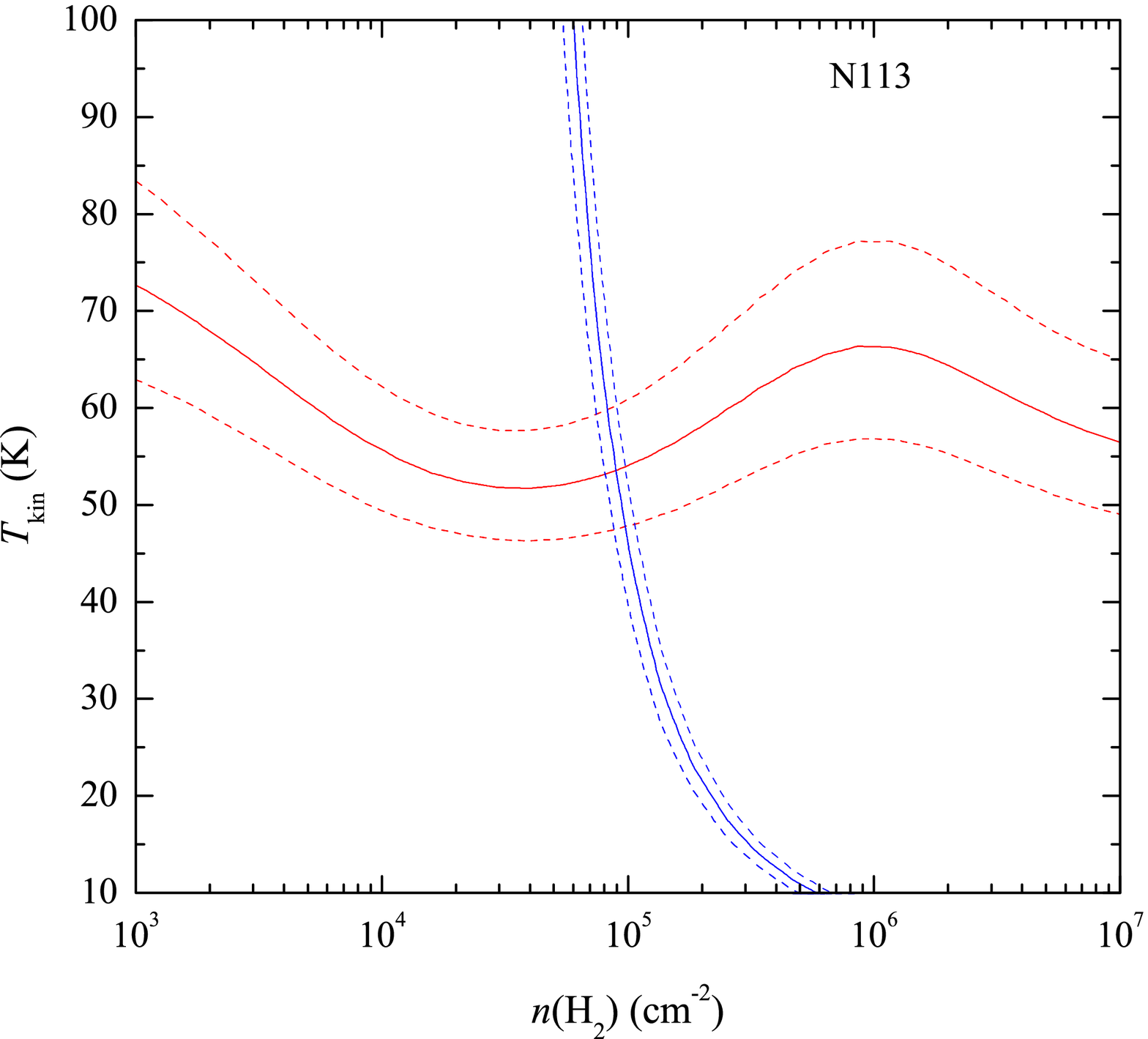}
\includegraphics[width=6cm]{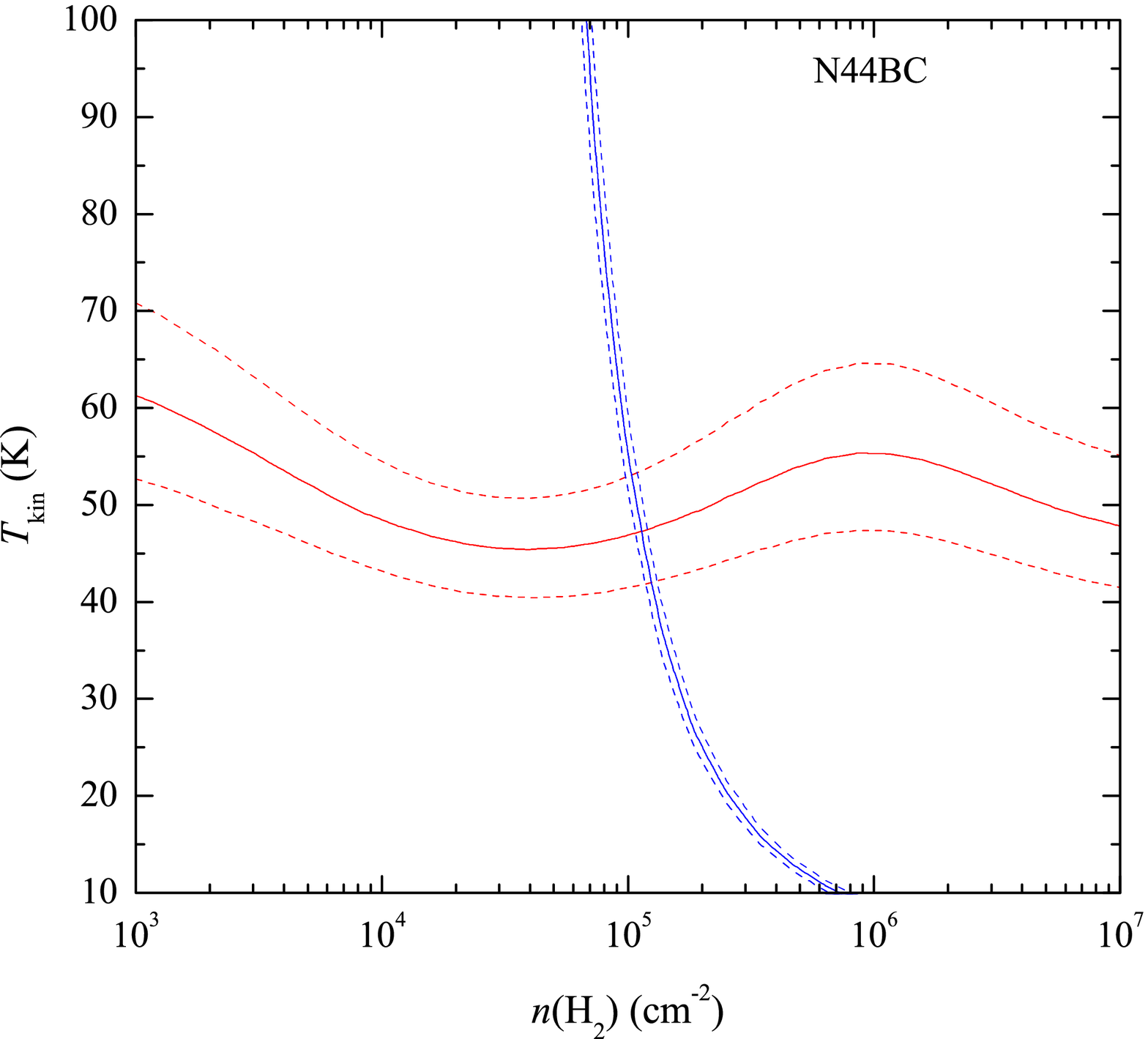}
\includegraphics[width=6cm]{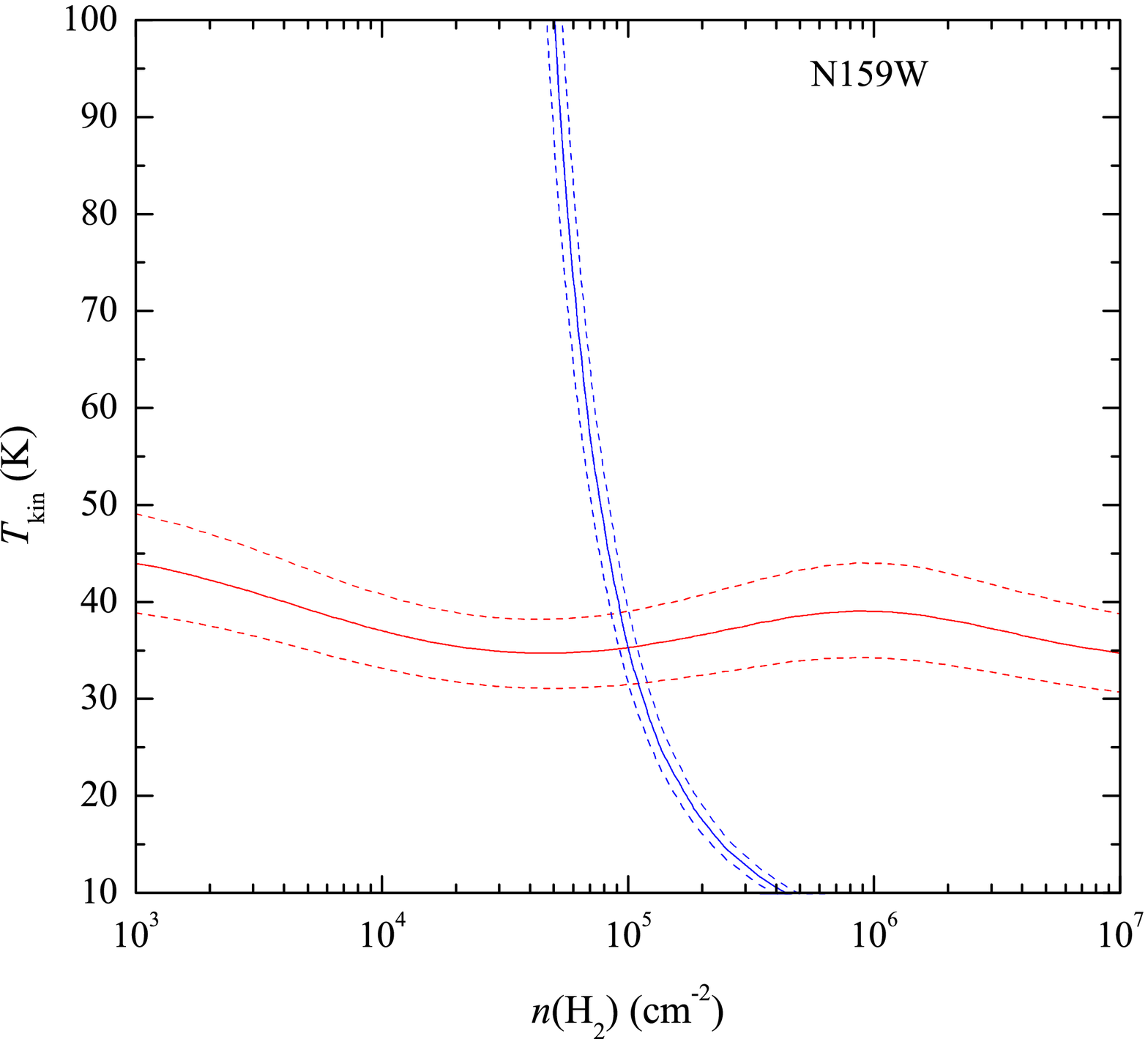}
\includegraphics[width=6cm]{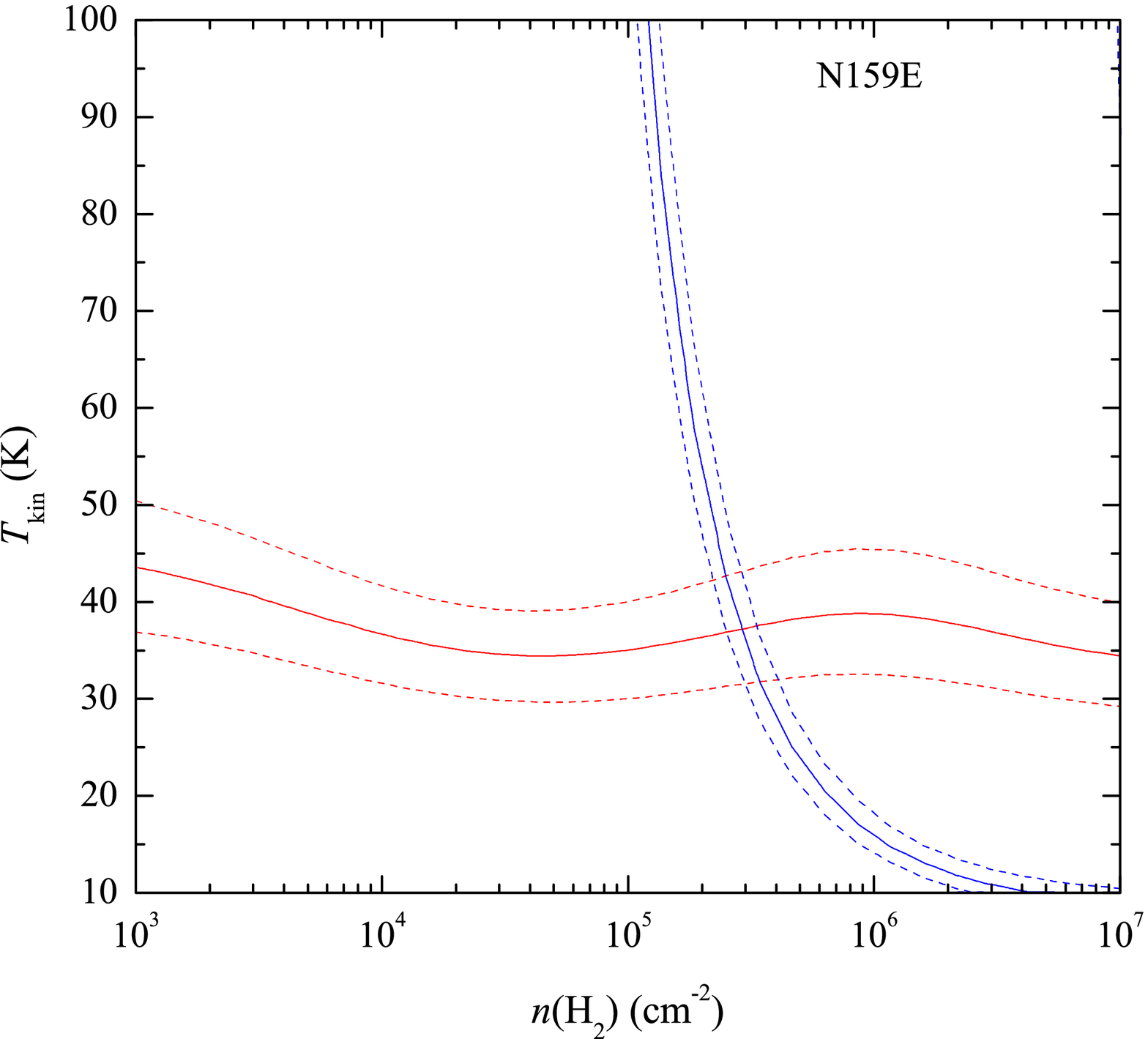}
\includegraphics[width=6cm]{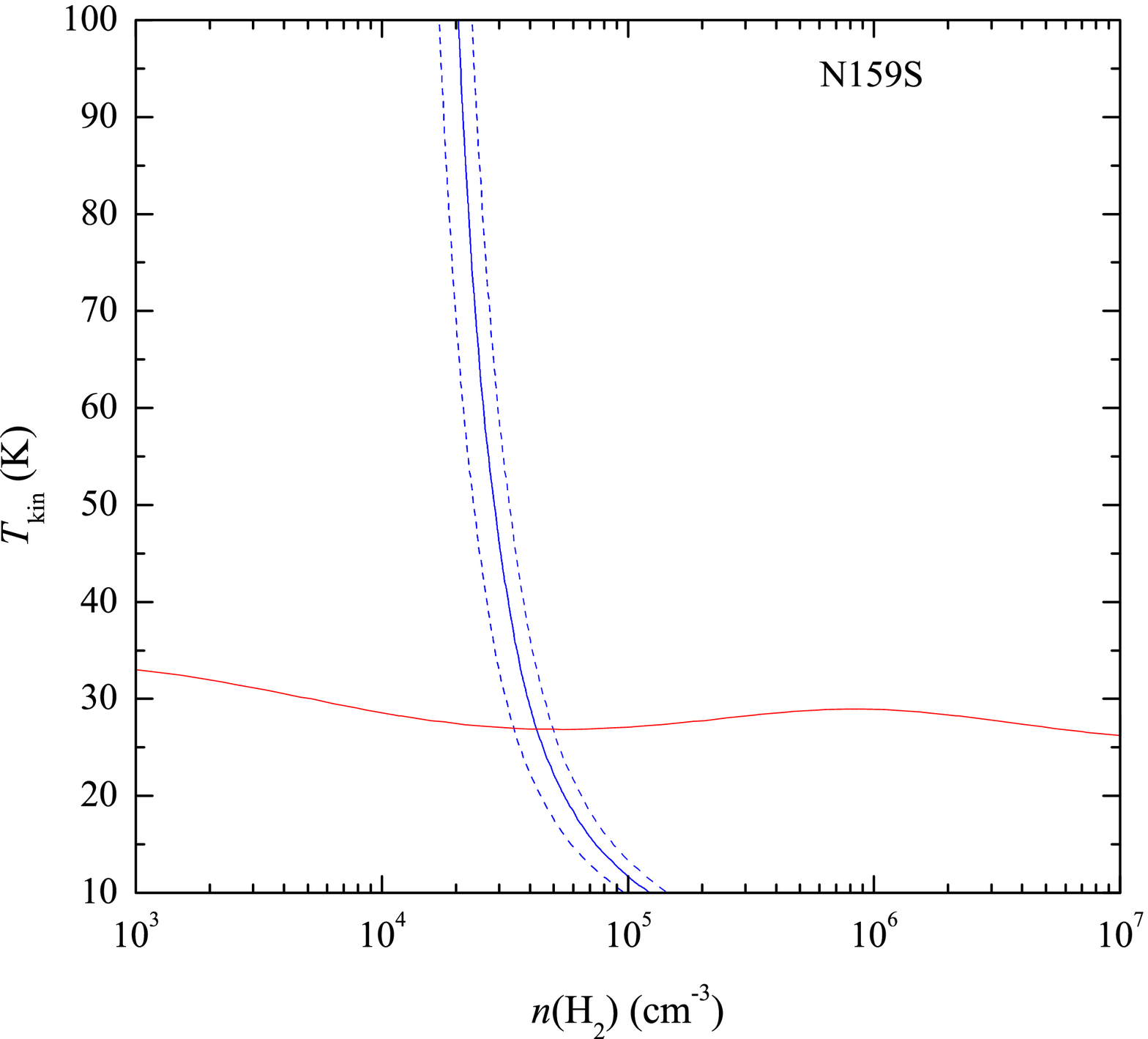}
\end{center}
\caption{RADEX non-LTE modeling of the gas kinetic temperature and
spatial density, using the para-H$_2$CO 3$_{21}$-2$_{20}$/3$_{03}$-2$_{02}$
(red solid and dashed lines are observed values and uncertainties)
and para-H$_2$CO 3$_{03}$-2$_{02}$/C$^{18}$O 2-1 (blue solid and
dashed lines) integrated intensity ratios.
For details, see the notes to Table \ref{table:Tkin}.}
\label{figure:Tkin-1}
\end{figure*}

\begin{table*}[t]
\small
\caption{Integrated intensity ratios, column densities, spatial
densities, and temperatures.}
\centering
\begin{tabular}
{ccccccccccccccccc}
\hline\hline 
&&&&&&\multicolumn{2}{c}{$N$(H$_2$)}&&&&\\
\cline{7-8}
Sources& 3$_{21}$-2$_{20}$ &&H$_2$CO 3$_{03}$-2$_{02}$ &$N$(C$^{18}$O)
&$N$(para-H$_2$CO) &C$^{18}$O &CO &$n$(H$_2$) &$T_{\rm kin}$ & $T_{\rm LTE}$
& $T_{\rm dust}$$^c$\\
\cline{2-2} \cline{4-4}
&3$_{03}$-2$_{02}$&&C$^{18}$O 2-1&$\times10^{14}$ cm$^{-2}$ &
$\times10^{12}$ cm$^{-2}$ &\multicolumn{2}{c}{$\times10^{22}$ cm$^{-2}$} &$\times10^{5}$ cm$^{-3}$ & K & K & K \\
\hline 
30 Dor    &0.27$\pm$0.05 &&2.12$\pm$0.32 &1.0  &1.0 &0.7 &0.6 &1.45$^{+0.69}_{-0.42}$ &63.0$^{+18.2}_{-13.9}$ &70.4 &70--75\\
N113$^a$  &0.23$\pm$0.03 &&1.22$\pm$0.12 &2.7  &2.7 &2.0 &1.0 &0.89$^{+0.18}_{-0.15}$ &53.6$^{+6.7}_{-6.1}$   &54.1 &30--51\\
N44BC     &0.20$\pm$0.03 &&1.39$\pm$0.07 &3.0  &3.0 &2.1 &1.3 &1.13$^{+0.18}_{-0.16}$ &47.3$^{+6.0}_{-5.4}$   &45.9 &35--45\\
N159W$^b$ &0.13$\pm$0.02 &&1.01$\pm$0.08 &3.7  &3.7 &2.7 &1.8 &1.00$^{+0.20}_{-0.14}$ &35.3$^{+3.7}_{-3.8}$   &32.0 &30--40\\
N159E     &0.13$\pm$0.03 &&2.39$\pm$0.21 &1.2  &1.2 &0.9 &0.8 &2.92$^{+1.15}_{-0.71}$ &37.2$^{+5.9}_{-5.7}$   &46.4 &30--40\\
N159S     &$<$0.085      &&0.35$\pm$0.07 &0.9  &0.9 &0.6 &1.4 &$>$0.43                &$<$26.9                &$<$25.1 &20--30\\
\hline 
\end{tabular}
\tablefoot{\\
Column 4: Obtained from equation (1).\\
Column 5: Para-H$_2$CO column densities adopting $N$(C$^{18}$O)/$N$(para-H$_2$CO) = 100 (see Section 3.1).\\
Column 6: H$_2$ column density derived from C$^{18}$O with
$X$(C$^{18}$O/H$_2$) = 4.2 $\times$ 10$^{-8}$ \citep{Frerking1982}
and using the factor 3 correction explained in Section 3.1.\\
Column 7: H$_2$ column density obtained from $^{12}$CO(1-0) with
a conversion factor of $X_{CO}$ = 4 $\times$ 10$^{20}$ cm$^{-2}$ (K km\,s$^{-1}$)$^{-1}$ (see Section 3.1).\\
Column 9: $T_{\rm kin}$ derived with the density given in column (8).\\
$^a$N113: Para-H$_2$CO 3$_{03}$-2$_{02}$/C$^{18}$O 2-1 ratio data
taken from \cite{Heikkila1998} and \cite{Wang2009}. $N$(C$^{18}$O)
derived from C$^{18}$O 2-1 taken from \cite{Heikkila1998}. \\
$^b$N159W: Para-H$_2$CO 3$_{03}$-2$_{02}$/C$^{18}$O 2-1 ratio data
taken from \cite{Heikkila1998,Heikkila1999}. $N$(C$^{18}$O) derived
from C$^{18}$O 2-1 taken from \cite{Heikkila1998}. $N$(H$_2$) data
taken from \cite{Ott2010}.\\
$^c$Dust temperature references: \\
30 Dor: \cite{Werner1978,Heikkila1999,Gordon2014}; \\
N113: \cite{Wang2009,Gordon2014,Seale2014}; \\
N44BC: \cite{Gordon2014}; \\
N159W: \cite{Heikkila1999,Bolatto2000,Gordon2014}; \\
N159E: \cite{Gordon2014};\\
N159S: \cite{Gordon2014}.\\
The upper and lower limits in columns 2, 8, 9, and 10 correspond to 3$\sigma$ values.}
\label{table:Tkin}
\end{table*}

The para-H$_2$CO (3$_{03}$-2$_{02}$) line is the strongest of the
three 218 GHz para-H$_2$CO transitions observed by us.
In order to avoid small uncertain values in the denominator, the
para-H$_2$CO 3$_{22}$-2$_{21}$/3$_{03}$-2$_{02}$ and 3$_{21}$-2$_{20}$/3$_{03}$-2$_{02}$
ratios are most suitable to derive the kinetic temperature.
The para-H$_2$CO 3$_{22}$-2$_{21}$ and 3$_{21}$-2$_{20}$
transitions have similar energy above the ground state,
$E_u$ $\simeq$ 68 K, similar line brightness, and are often detected
at the same time (e.g.,
\citealt{Muhle2007,Bergman2011,Wang2012,Lindberg2012,Ao2013,Immer2014,
Immer2016,Trevino2014,Ginsburg2016,Tang2016});
therefore, para-H$_2$CO 3$_{22}$-2$_{21}$/3$_{03}$-2$_{02}$ and
3$_{21}$-2$_{20}$/3$_{03}$-2$_{02}$ ratios are both good thermometers
to determine the gas temperature. The kinetic temperature is
traced by these two ratios with an uncertainty of $\lesssim$ 25\%
below 50 K \citep{Mangum1993a}.
While at $n$(H$_2$) $\gtrsim$ 10$^5$ cm$^{-3}$ both ratios are similarly suitable,
the para-H$_2$CO 3$_{22}$-2$_{21}$/3$_{03}$-2$_{02}$ line ratio
is affected by gas density at
$n$(H$_2$) $<$ 10$^5$ cm$^{-3}$ \citep{Tang2016}. For our sample,
spatial density measurements probed by molecular tracers like CS, SO,
CO, CI, H$_2$CO, HCO$^+$, and HCN \citep{Heikkila1998,Heikkila1999,Kim2004,Wang2009}
show a range of 0.3 -- 10 $\times$ 10$^5$ cm$^{-3}$. Therefore, in this work
we use the para-H$_2$CO 3$_{21}$-2$_{20}$/3$_{03}$-2$_{02}$ integrated intensity
ratio to derive the kinetic temperature \citep{Ginsburg2016,Immer2016}.


In our Galaxy, spatial densities, $n$(H$_2$), derived from
para-H$_2$CO (3$_{03}$-2$_{02}$) are higher than from C$^{18}$O (2-1).
However, in the LMC, with lower density regions often being photoionized,
this may be different. Therefore, here the
para-H$_2$CO 3$_{03}$-2$_{02}$/C$^{18}$O 2-1 integrated intensity
ratio is used to constrain the spatial density assuming the two
tracers have a similar spatial extent and sample the same region. For both lines
we find similar line parameters (e.g., $V_{\rm lsr}$, FWHM,
see Table \ref{table:H2CO-C18O}) in our sample, so our assumption is reasonable.
igure \ref{figure:Tkin-1} shows how the parameters are constrained by
the line ratio distribution of
para-H$_2$CO 3$_{21}$-2$_{20}$/3$_{03}$-2$_{02}$ and
para-H$_2$CO 3$_{03}$-2$_{02}$/C$^{18}$O 2-1 in the
$T_{\rm kin}$-$n$(H$_2$) parameter space. The determined
results are listed in Table \ref{table:Tkin}.
The spatial density of our sample derived from the
para-H$_2$CO 3$_{03}$-2$_{02}$/C$^{18}$O 2-1 ratio shows a relatively
narrow range of 0.4 -- 2.9 $\times$ 10$^5$ cm$^{-3}$ with an average of
1.5 $\pm$ 0.4 $\times$ 10$^5$ cm$^{-3}$ (see Figure \ref{figure:Tkin-1}
and Table \ref{table:Tkin}; errors given here and elsewhere are unweighted
standard deviations of the mean), which is consistent with the results
for the same dense clumps found from observations of e.g., SO,
HCO$^+$, c-C$_3$H$_2$, CH$_3$OH, and H$_2$CO \citep{Heikkila1999,Wang2009}.

\begin{figure}[t]
\vspace*{0.2mm}
\begin{center}
\includegraphics[width=9cm]{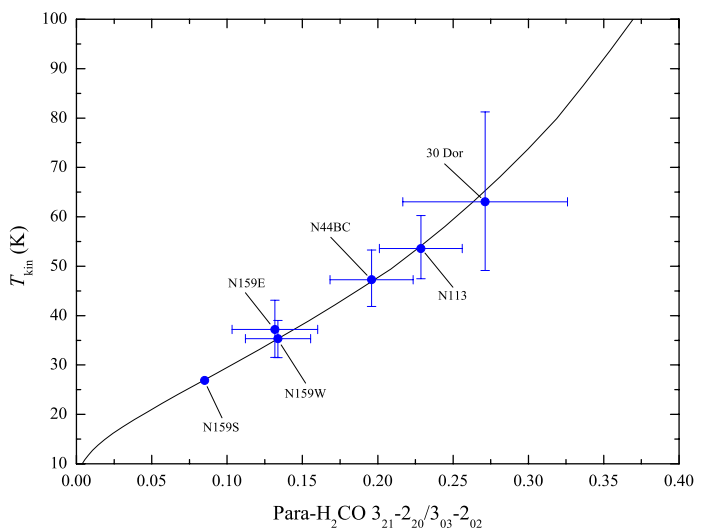}
\end{center}
\caption{RADEX non-LTE modeling of the relation between the gas
kinetic temperature and the para-H$_2$CO 3$_{21}$-2$_{20}$/3$_{03}$-2$_{02}$
integrated intensity ratio (solid line), assuming
$n$(H$_2$) = 10$^{5}$ cm$^{-3}$,
$N$(para-H$_2$CO) = 2.1 $\times$ 10$^{12}$ cm$^{-2}$, and a line
width of 6 km\,s$^{-1}$. Blue points are
the results derived from the measured para-H$_2$CO 3$_{21}$-2$_{20}$/3$_{03}$-2$_{02}$
and para-H$_2$CO 3$_{03}$-2$_{02}$/C$^{18}$O 2-1 ratios.
For N159S, 3$\sigma$ upper limits for the line ratio and
the kinetic temperature have been taken.}
\label{figure:Tk-ratio-n-1e5}
\end{figure}

As already mentioned, our three para-H$_2$CO (3-2) transitions
are sensitive to gas at density
$\sim$10$^5$ cm$^{-3}$ \citep{Immer2016}.
To highlight how much the derived kinetic temperature depends on the derived density
based on the admittedly uncertain assumption that
para-H$_2$CO 3$_{03}$-2$_{02}$ and C$^{18}$O 2-1 trace the same gas,
we have plotted  in Figure \ref{figure:Tk-ratio-n-1e5}
the relation between kinetic temperature and
the para-H$_2$CO 3$_{21}$-2$_{20}$/3$_{03}$-2$_{02}$ ratio at
spatial density $n$(H$_2$) = 10$^{5}$ cm$^{-3}$ with
column density of $N$(para-H$_2$CO) = 2.1 $\times$ 10$^{12}$ cm$^{-2}$
and line width of 6 km\,s$^{-1}$ (these are rough average values for our sample)
using RADEX. Comparing this with the actually obtained values
for individual sources derived from our
para-H$_2$CO 3$_{21}$-2$_{20}$/3$_{03}$-2$_{02}$ and
para-H$_2$CO 3$_{03}$-2$_{02}$/C$^{18}$O 2-1 ratios,
the plot demonstrates that the temperatures derived
in the two different ways are in a good agreement.

Local thermodynamic equilibrium (LTE) is a good approximation
for the H$_2$CO level populations under optically thin
high-density conditions \citep{Mangum1993a,Watanabe2008}.
Although LTE and RADEX non-LTE models use different
approximations and assumptions, it is useful to check how
the temperatures derived by the two methods compare.
The para-H$_2$CO line intensity ratios
3$_{22}$-2$_{21}$/3$_{03}$-2$_{02}$ and
3$_{21}$-2$_{20}$/3$_{03}$-2$_{02}$ can be used to measure
the LTE kinetic temperature because (see Section 1) the $K_{\rm a}$ = 0 and 2 ladders
of para-H$_2$CO are mainly connected by collisions.
The LTE kinetic temperature can be calculated
assuming that the lines are optically thin, and originate
from a high-density region \citep{Mangum1993a}.
\begin{equation}
T_{\rm kin} =  \frac{47.1}{ln(0.556\frac{I(3_{03}-2_{02})}{I(3_{22}-2_{21})})},
\end{equation}
where $I$(3$_{03}$-2$_{02}$)/$I$(3$_{22}$-2$_{21}$) is the
para-H$_2$CO integrated intensity ratio. The LTE kinetic temperatures
are listed in Table \ref{table:Tkin}. The uncertainty is
$\lesssim$ 30\% for this method of temperature measurement
\citep{Mangum1993a}. Considering this uncertainty, the temperatures
derived from LTE and the RADEX non-LTE model are consistent with each other.

\subsection{Individual sources}
Below, results from the six sources covered by this study are
individually discussed.

\subsubsection{30 Dor}
Our three para-H$_2$CO and C$^{18}$O 2-1 transitions have
already been observed toward 30 Dor by \cite{Heikkila1999} with
the SEST (beam size $\sim$23$''$), but only
para-H$_2$CO 3$_{03}$-2$_{02}$ was detected. With our higher
sensitivity we detect all four lines and confirm their
para-H$_2$CO results. The kinetic temperature derived from
para-H$_2$CO (3$_{21}$-2$_{20}$/3$_{03}$-2$_{02}$) is $\sim$63 K,
which is the highest
value determined in our sample.
30 Dor, hosting a cluster of O3 stars that rivals super
star clusters, is the most spectacular star forming region
in the Local Group \citep{Walborn1997,Massey1998},
which makes such a high $T_{\rm kin}$ value comprehensible.
The spatial density derived from the
para-H$_2$CO 3$_{03}$-2$_{02}$/C$^{18}$O 2-1 ratio is
$\sim$1.5 $\times$ 10$^5$ cm$^{-3}$ at this temperature,
which agrees with results derived from other species
(e.g., CS, SO, HCO$^+$, \citealt{Heikkila1999}).

\subsubsection{N113}
All three 218 GHz para-H$_2$CO lines as well as
C$^{18}$O 2-1 transition have been observed by \cite{Wang2009} and
\cite{Heikkila1998} with the SEST. Para-H$_2$CO 3$_{03}$-2$_{02}$
and C$^{18}$O 2-1 were detected. We detect the three transitions
of para-H$_2$CO for the first time. The kinetic temperature
derived from para-H$_2$CO (3$_{21}$-2$_{20}$/3$_{03}$-2$_{02}$) is $\sim$54 K.
N113 is the LMC star forming region with the most luminous 22 GHz
H$_2$O maser \citep{Whiteoak1986,Lazendic2002,Oliveira2006},
not quite as spectacular as 30 Dor, but hosts a few bright HII regions whose stellar
energy feedback is likely to have elevated its temperature,
thus leading to the second highest $T_{\rm kin}$ value. The spatial
density derived from para-H$_2$CO 3$_{03}$-2$_{02}$/C$^{18}$O 2-1
(ratio data from \cite{Wang2009} and \cite{Heikkila1998}, assuming
that the para-H$_2$CO 3$_{03}$-2$_{02}$/C$^{18}$O 2-1 ratio
at the SEST beam size, $\sim$23$''$, is similar to that in the APEX
beam size, $\sim$30$''$, see Section 3.2.4) is
$\sim$8.9 $\times$ 10$^4$ cm$^{-3}$ at temperature $\sim$54 K,
which agrees with results from other molecules (e.g., CS, HCO$^+$, HCN,
\citealt{Wang2009}).

\subsubsection{N44BC}
C$^{18}$O 2-1 has been detected in N44BC by \cite{Heikkila1998}
with the SEST. We detect the three 218 GHz para-H$_2$CO transitions and
the C$^{18}$O line. Our observations confirm their C$^{18}$O 2-1 result.
The kinetic temperature derived by
para-H$_2$CO (3$_{21}$-2$_{20}$/3$_{03}$-2$_{02}$) is $\sim$47 K.
Such a high temperature likely results from the stellar energy
feedback from the adjacent super bubble on the molecular cloud,
which also shows bright mid-IR emission \citep{Chen2009}.
The spatial density derived from the para-H$_2$CO 3$_{03}$-2$_{02}$/C$^{18}$O 2-1
ratio is $\sim$1.1 $\times$ 10$^5$ cm$^{-3}$ at this temperature,
which agrees with results from other molecules
(e.g., CO, CI, \citealt{Heikkila1998,Kim2004}).

\subsubsection{N159W}
The three para-H$_2$CO 218 GHz transitions  as well as C$^{18}$O 2-1
have been detected in N159W by \cite{Heikkila1998,Heikkila1999}
with the SEST. Our observations confirm their para-H$_2$CO results.
The kinetic temperature derived by
para-H$_2$CO (3$_{21}$-2$_{20}$/3$_{03}$-2$_{02}$) is $\sim$35 K.
The spatial density
derived from the para-H$_2$CO 3$_{03}$-2$_{02}$/C$^{18}$O 2-1
ratio (data from \citealt{Heikkila1998,Heikkila1999}) is
$\sim$1.0 $\times$ 10$^5$ cm$^{-3}$ at this temperature,
which agrees with results measured by other species (e.g., CS, SO,
\citealt{Heikkila1999}). To quantify potential differences in
temperature and density derived from APEX and SEST data, we determined
the temperature and the density with the same method using the
SEST para-H$_2$CO and C$^{18}$O data from \cite{Heikkila1998,Heikkila1999}.
The derived temperature and density are 30.3$^{+5.7}_{-5.4}$ K and
1.14$^{+0.35}_{-0.23}$ $\times$ 10$^5$ cm$^{-3}$, respectively.
This indicates nearly the same spatial density
and a few Kelvin difference for the kinetic temperature. This
temperature difference is similar to its 1$\sigma$ uncertainty
($\sim$6 K). Therefore, the temperature gradient and density gradient
is small when moving from a beam size of 7.3 pc (APEX)
to 5.6 pc (SEST).

\subsubsection{N159E}
The three para-H$_2$CO 218 GHz transitions as well as
C$^{18}$O 2-1 are detected. To our knowledge, it is the first
detection of para-H$_2$CO in the N159E region. The source shows
similar properties as N159W. The kinetic temperature derived from
para-H$_2$CO (3$_{21}$-2$_{20}$/3$_{03}$-2$_{02}$) is $\sim$37 K.
The spatial density derived from the para-H$_2$CO 3$_{03}$-2$_{02}$/C$^{18}$O 2-1
ratio is $\sim$2.9 $\times$ 10$^5$ cm$^{-3}$ at this temperature.

\subsubsection{N159S}
Two velocity components are detected by C$^{18}$O, at 232.4
and 236.4 km\,s$^{-1}$. Para-H$_2$CO 3$_{03}$-2$_{02}$ is detected
at a velocity of 235.9 km\,s$^{-1}$ (see Table 2). However, the
para-H$_2$CO 3$_{22}$-2$_{21}$ and 3$_{21}$-2$_{20}$ lines are
not detected. The dust temperature ranges from 20 to 30 K \citep{Gordon2014}.
N159S appears to be a cold cloud \citep{Heikkila1999},
and has been shown to host no massive star formation at present
and during the last 10 Myr \citep{Chen2010}. The upper limit
to the kinetic temperature derived from
para-H$_2$CO (3$_{21}$-2$_{20}$/3$_{03}$-2$_{02}$)
based on our observational 3$\sigma$ limit for the
3$_{21}$-2$_{20}$ line is $\sim$27 K.
The spatial density derived from the
para-H$_2$CO 3$_{03}$-2$_{02}$/C$^{18}$O 2-1 ratio is
$>$4.3 $\times$ 10$^4$ cm$^{-3}$, which is consistent with
results measured by other species (e.g., CS, SO, \citealt{Heikkila1999}).

\section{Discussion}
\subsection{Comparison of temperatures derived from H$_2$CO, CO,
NH$_3$, and the dust}
The kinetic temperatures of molecular clumps in the LMC have
been calculated by multi-transition data of CO
\citep{Johansson1998,Heikkila1999,Israel2003,Kim2004,Bolatto2005,
Pineda2008,Mizuno2010,Minamidani2008,Minamidani2011,Fujii2014,Paron2016}.
These observations show that the higher temperatures
($T_{\rm kin}$ $\gtrsim$ 100 K) occur in cloud regions that
are of lower density ($\lesssim$10$^3$ cm$^{-3}$) and that
the gas is colder ($T_{\rm kin}$ = 10 -- 80 K) in regions of
higher density (10$^4$ -- 10$^5$ cm$^{-3}$). As already
mentioned in Section 3.1, the spatial density range
of our sample derived from the
para-H$_2$CO 3$_{03}$-2$_{02}$/C$^{18}$O 2-1 ratio with respect to
the Galaxy is 0.4 -- 2.9 $\times$ 10$^5$ cm$^{-3}$ with
an average of 1.5 $\pm$ 0.4 $\times$ 10$^5$ cm$^{-3}$.
Excluding the quiescent cloud N159S, where only one
para-H$_2$CO line could be detected, the gas kinetic
temperatures derived from para-H$_2$CO (3$_{21}$-2$_{20}$/3$_{03}$-2$_{02}$),
range from 35 to 63 K with an average of 47 $\pm$ 5 K. Temperatures
and densities derived from CO are for 30 Dor $T_{\rm kin}$
$\sim$ 40 -- 80 K and $n$(H$_2$) $\sim$ 3$\times$10$^3$ -- 3$\times$10$^5$ cm$^{-3}$
\citep{Johansson1998,Heikkila1999,Minamidani2008},
N159W $T_{\rm kin}$ $\sim$ 16 -- $>$30 K and
$n$(H$_2$) $\sim$ 3$\times$10$^3$ -- 8$\times$10$^5$ cm$^{-3}$
\citep{Johansson1998,Heikkila1999,Bolatto2005,Minamidani2008},
N159E $T_{\rm kin}$ $>$ 40 K and
$n$(H$_2$) $\sim$ 1$\times$10$^3$ -- 3$\times$10$^5$ cm$^{-3}$
\citep{Minamidani2008}, and N159S $T_{\rm kin}$ $\sim$ 10 -- 60 K
and $n$(H$_2$) $\sim$ 1$\times$10$^3$ -- 1$\times$10$^5$ cm$^{-3}$
\citep{Heikkila1999,Minamidani2008}. Temperatures derived
from para-H$_2$CO are consistent with but much more precise
than the results derived from CO in the dense regions
($>$10$^3$ cm$^{-3}$).

\begin{figure}[t]
\vspace*{0.2mm}
\begin{center}
\includegraphics[width=9cm]{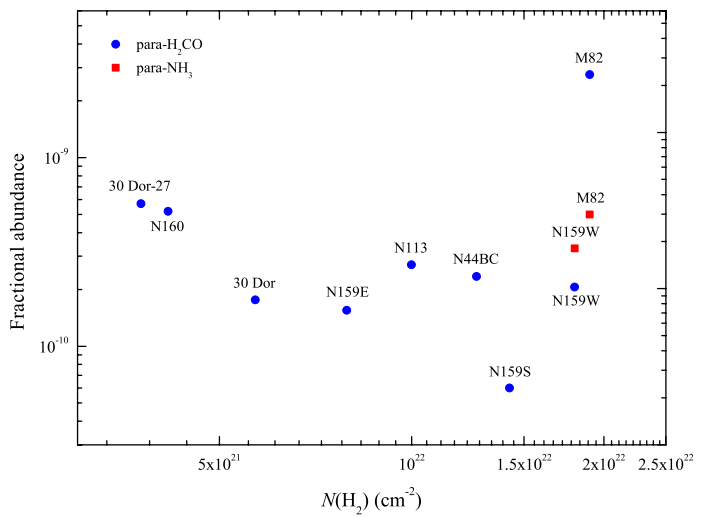}
\end{center}
\caption{Abundances of $N$(para-NH$_3$)/$N$(H$_2$), red squares,
and $N$(para-H$_2$CO)/$N$(H$_2$), blue points, vs. column
density $N$(H$_2$) obtained from $^{12}$CO. 30 Dor-27 and N160 data taken from
\cite{Heikkila1999}. N159W data taken from \cite{Ott2010}.
M82 data taken from \cite{Weiss2001} and \cite{Muhle2007}.}
\label{figure:X(H2CO)-X(NH3)}
\end{figure}

Except for N159E, all our sources have been surveyed in NH$_3$ (1,1)
and (2,2) by \cite{Ott2010}. These lines are only detected in
the massive star-forming region N159W. We compare fractional
abundances of $N$(para-NH$_3$)/$N$(H$_2$) and
$N$(para-H$_2$CO)/$N$(H$_2$) to column density $N$(H$_2$)
obtained from $^{12}$CO(1-0) in Figure \ref{figure:X(H2CO)-X(NH3)}.
These show that formaldehyde
has a stable fractional abundance ranging from 0.6 to
5.7 $\times$ 10$^{-10}$ cm$^{-2}$ with an average of
2.7 $\pm$ 1.8 $\times$ 10$^{-10}$ cm$^{-2}$ in molecular clouds
of the LMC with $N$(H$_2$) column densities ranging from 0.4 to
2 $\times$ 10$^{22}$ cm$^{-2}$. Ammonia only survives in a
high column density environment with
$N$(H$_2$) $\sim$ 2 $\times$ 10$^{22}$ cm$^{-2}$. The fractional
abundance of ammonia is $\sim$ 10$^{-10}$ -- 10$^{-9}$ in N159W and M82,
which is similar to that of formaldehyde in the LMC. As already
mentioned, the kinetic temperature derived from the NH$_3$ (2,2)/(1,1)
line ratio in N159W is $\sim$16 K \citep{Ott2010}, which is two times lower than that
derived from para-H$_2$CO. Previous
para-H$_2$CO (3$_{22}$-2$_{21}$/3$_{03}$-2$_{02}$) and NH$_3$ (2,2)/(1,1)
observations toward the starburst galaxy M82 also show
significantly different gas kinetic temperatures \citep{Weiss2001,Muhle2007}.
M82, a satellite galaxy like the LMC, shows a similar environment,
involving low metallicity combined with a high UV flux. This only
leaves NH$_3$ surviving in the most shielded pockets of molecular gas,
resulting in a low fractional abundance (see Figure \ref{figure:X(H2CO)-X(NH3)})
and a low kinetic temperature. Furthermore, this abundance is
demonstrating in an exemplary way that H$_2$CO is less affected by
photodissociation, sampling a more extended region.
Therefore, para-H$_2$CO traces in these instances a higher temperature
than NH$_3$ (2,2)/(1,1). We conclude that para-H$_2$CO line ratios
are a superior thermometer to trace dense gas temperatures in low
metallicity galaxies with strong UV radiation. Nevertheless,
more detailed spatially resolved comparisons of H$_2$CO with NH$_3$
temperatures would be very interesting because it could provide more
information on temperature gradients and the location of different
kinetic temperature layers.

The temperatures derived from dust and gas are often in agreement
in the active and dense clumps of Galactic disk clouds
\citep{Dunham2010,Giannetti2013,Battersby2014}. However, observed
gas and dust temperatures do not agree with each other in the Galactic Central
Molecular Zone (CMZ) and external galaxies
\citep{Gusten1981,Ao2013,Mangum2013a,Ott2014,Ginsburg2016,Immer2016}.
Dust temperatures in the LMC have been obtained by \cite{Gordon2014}
using Herschel 100 to 500 $\mu$m dust continuum emission data.
They range approximately from 13 to 73 K. For our sample, the dust
temperatures range from 30 to 75 K
\citep{Werner1978,Heikkila1999,Bolatto2000,Wang2009,Seale2014,Gordon2014}
while the para-H$_2$CO derived temperature
ranges from 35 to 63 K. The temperatures
derived from para-H$_2$CO ratios and dust emission are therefore in
good agreement (see Table \ref{table:Tkin}). This indicates that the
dust and H$_2$CO kinetic temperatures are equivalent in the star
forming regions of the LMC. Assuming that H$_2$CO traces the bulk
of the dense molecular gas and that ammonia shows very low abundances,
this can be generalized in the sense that dense gas and dust temperatures
are generally equivalent.

\subsection{Star-forming regions in the Galaxy, the LMC, and other galaxies}
The gas temperatures of ATLASGAL (APEX Telescope Large Area Survey
of the GALaxy) massive star forming clumps have been measured by
para-H$_2$CO (3$_{03}$-2$_{02}$, 3$_{22}$-2$_{21}$, and 3$_{21}$-2$_{20}$)
line ratios \citep{Tang2016}. The thus derived gas kinetic temperatures
at density n(H$_2$) = 10$^5$ cm$^{-3}$
with size of 0.3 -- 0.7 pc range from 30 to 61 K with an average of
46 $\pm$ 9 K, which agrees remarkably well with the results in the
LMC with a beam size of $\sim$7 pc. Large area mapping measurements
of kinetic temperatures in the Galactic CMZ with the same transitions
of para-H$_2$CO \citep{Ao2013,Ginsburg2016} suggest that the mean gas
temperature is $\sim$48 K at $n$(H$_2$) = 10$^5$ cm$^{-3}$ or $\sim$65 K
at $n$(H$_2$) = 10$^4$ cm$^{-3}$ in the whole $\sim$300 pc surveyed region.
It shows a higher $T_{\rm kin}$ value than our observed results. The spatial
densities derived from the para-H$_2$CO 3$_{03}$-2$_{02}$/C$^{18}$O 2-1
ratio in our sample agree with the observed results in the Galactic
clumps \citep{Beuther2002,Motte2003,Wienen2012,Wienen2015}, HII regions
\citep{Henkel1983,Ginsburg2011}, and giant molecular clouds (GMCs)
\citep{Wadiak1988,Ginsburg2015,Immer2016}. This agreement indicates
that the physical conditions of the star forming regions should be
similar in both the LMC and our Galactic disk.

Using the three transitions of para-H$_2$CO at $\sim$218 GHz to measure
the kinetic temperature of the starburst galaxy M82 shows that the
derived kinetic temperature ($T_{\rm kin}$(H$_2$CO) $\sim$ 200 K;
\citealt{Muhle2007}) is significantly higher than in the LMC.
Kinetic temperatures of starburst galaxies
measured with multi-inversion transitions of NH$_3$ show a range
from 24 to $\gtrsim$ 250 K
\citep{Henkel2000,Henkel2008,Mauersberger2003,Ao2011,Lebron2011,Mangum2013a}.
The temperatures derived from para-H$_2$CO line ratios in the LMC
overlap with the values found for a sample at lower temperature (e.g., M83, NGC6946).
This is likely due to the inclusion of higher
excited ammonia lines, which, however, should be difficult to detect
in the LMC because there particularly warm regions irradiated
by enhanced UV radiation should be almost devoid of NH$_3$.
The spatial densities in starburst galaxies
derived from ortho-H$_2$CO (2$_{11}$-2$_{12}$/1$_{10}$-1$_{11}$)
line ratios \citep{Mangum2008,Mangum2013b} show a similar range
(10$^{4.5}$--10$^{5.5}$ cm$^{-3}$) to that derived from
para-H$_2$CO 3$_{03}$-2$_{02}$/C$^{18}$O 2-1 ratios in the LMC.
This indicates that star formation in the LMC and external galaxies
may arise from dense molecular gas ($>$10$^4$ cm$^{-3}$),
but gas heating rates may be quite different.

\subsection{Correlation of gas temperature with star formation}
\begin{figure}[t]
\vspace*{0.2mm}
\begin{center}
\includegraphics[width=9cm]{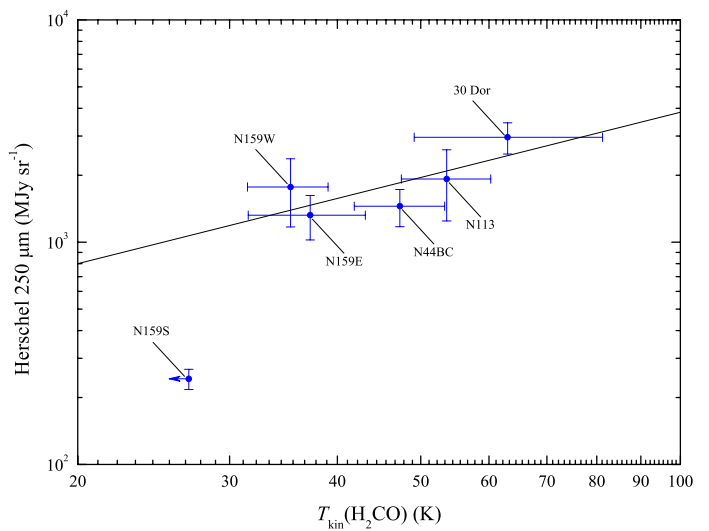}
\end{center}
\caption{Comparison of gas kinetic temperatures derived from
para-H$_2$CO 3$_{21}$-2$_{20}$/3$_{03}$-2$_{02}$ against the
Herschel 250 $\mu$m flux. The straight line is the result from
a linear fit for the five sources with all 218 GHz para-H$_2$CO lines detected.
N159S provides with respect to $T_{\rm kin}$ a 3$\sigma$ upper limit.
Thus this point may actually be located much further to the left,
closer to the linear fit obtained from the other five sources.}
\label{figure:Tk-F250um}
\end{figure}

In star forming galaxies, a lack of correlation between the gas
kinetic temperatures traced by NH$_3$ and star formation rate
indicated by infrared luminosity was found by \cite{Mangum2013a}.
To investigate how the kinetic temperatures traced by para-H$_2$CO
correlate with star formation in the LMC, we compared the gas
kinetic temperature to the Herschel 250 $\mu$m dust emission which
indicates the infrared luminosity \citep{Seale2014}. We averaged the
Herschel 250 $\mu$m data over a 30$''$ beam corresponding to our
para-H$_2$CO data. A comparison between gas kinetic temperatures
derived from para-H$_2$CO and the Herschel 250 $\mu$m flux is shown
in Figure \ref{figure:Tk-F250um}. It reveals a correlation of gas
temperature and 250 $\mu$m flux for the five sources with all
218 GHz para-H$_2$CO lines detected (slope = 0.97 $\pm$ 0.49,
correlation coefficient
$R$ $\sim$ 0.75). According to the relation between the infrared
luminosity and 250 $\mu$m flux $L_{\rm FIR}$ $\propto$ $F_{\rm S250}^{1.2}$
in the LMC \citep{Seale2014}, the infrared
luminosity and the gas temperature derived from para-H$_2$CO are related
by a power-law of the form $L_{\rm FIR}$ $\propto$ $T_{\rm kin}^{1.2\pm0.59}$
where the power-law index is lower than that of the Stefan-Boltzmann law
($L$ $\propto$ $T_{\rm kin}^4$). This suggests that this picture is an
oversimplification, and that star formation occurs in extended regions
leading to the formation of stellar clusters with multiple FIR sources
(e.g., \citealt{Chen2009,Chen2010}).
Assuming that the 250 $\mu$m flux is mostly coming from clusters of
FIR sources distributed across the regions from where the H$_2$CO
emission is arising, this can be generalized in the sense that the
gas heating mechanism must be related to the formation of young massive stars.
To find out whether the bulk of the H$_2$CO emission is originating from
within individual clusters of FIR sources, in between adjacent such
FIR clusters, or both, higher resolution H$_2$CO observations would be mandatory.

We need more data points in the LMC to understand the relationship
between $T_{\rm kin}$ and star formation and to then apply this
relationship to more distant galaxies with ALMA.
Para-H$_2$CO 3$_{21}$-2$_{20}$/3$_{03}$-2$_{02}$ and
para-H$_2$CO 3$_{03}$-2$_{02}$/C$^{18}$O 2-1 line ratios provide
a direct estimate of the gas kinetic temperatures and spatial
densities for molecular gas on a scale of $\sim$7 pc in the
star forming regions of the LMC. It would be meaningful to use
these line ratios to measure the physical properties of the dense
molecular gas at smaller linear scales with ALMA and to start
systematic investigations in more distant star-forming galaxies.

\section{Summary}
We have measured the kinetic temperature and spatial density with
para-H$_2$CO ($J$$_{K_AK_C}$ = 3$_{03}$-2$_{02}$, 3$_{22}$-2$_{21}$,
and 3$_{21}$-2$_{20}$) line ratios and the C$^{18}$O 2-1 line
in massive star-forming regions of the LMC. Kinetic temperatures
derived from the above mentioned formaldehyde 218 GHz line triplet
are compared with those obtained from the dust and, in one case,
also from ammonia using the 12-m APEX telescope.
The main results are the following:
\begin{enumerate}
\item
Using the RADEX non-LTE program, we derive gas kinetic temperatures
and spatial densities modeling the measured
para-H$_2$CO 3$_{21}$-2$_{20}$/3$_{03}$-2$_{02}$ and
para-H$_2$CO 3$_{03}$-2$_{02}$/C$^{18}$O 2-1 line ratios.

\item
The gas kinetic temperatures derived from
para-H$_2$CO (3$_{21}$-2$_{20}$/3$_{03}$-2$_{02}$) line ratios
of the star forming
regions in the LMC are warm, ranging from 35 to 63 K with an average
of 47 $\pm$ 5 K, which is similar to that
obtained from Galactic disk massive star forming clumps.

\item
The spatial density derived from the
para-H$_2$CO 3$_{03}$-2$_{02}$/C$^{18}$O 2-1 ratio shows
a range of 0.4 -- 2.9 $\times$ 10$^5$ cm$^{-3}$
with an average of 1.5 $\pm$ 0.4 $\times$ 10$^5$ cm$^{-3}$.
It agrees with results measured by
ortho-H$_2$CO (2$_{11}$-2$_{12}$/1$_{10}$-1$_{11}$) line ratios in
Galactic regions of massive star formation.

\item
Temperatures derived from the para-H$_2$CO line ratios agree
with those derived from CO in dense regions ($n$(H$_2$) $>$ 10$^3$ cm$^{-3}$).
The gas temperature derived from the NH$_3$ (2,2)/(1,1)
line ratio is $\sim$16 K in N159W \citep{Ott2010}, which is
two times lower than the temperature derived from the para-H$_2$CO
line ratio and the dust. Ammonia only survives in the most
shielded pockets of molecular gas in the LMC$'$s low metallicity environment
affected by a high UV flux. Formaldehyde is less affected by
photodissociation and traces a more extended region.

\item
A comparison of the gas kinetic temperature derived from
para-H$_2$CO and the temperature obtained from dust emission
shows good agreement.
It indicates that the bulk of the dense gas and dust are in approximate thermal
equilibrium in the dense star formation regions of the LMC.

\item
A correlation between the gas kinetic temperatures
derived from para-H$_2$CO and infrared luminosity indicated by
the 250 $\mu$m flux suggests that our kinetic temperatures traced
by para-H$_2$CO are closely associated with extended star formation
in the LMC.
\end{enumerate}

\begin{acknowledgements}
We thank the staff of the APEX telescope for their assistance in
observations. The authors are also thankful for the helpful comments
of the anonymous referee. This work was funded by The National Natural Science
Foundation of China under grant 11433008 and The Program of the
Light in China$'$s Western Region (LCRW) under grant Nos.XBBS201424
and The National Natural Science Foundation of China under grant 11373062.
C.H acknowledges support by a visiting professorship for senior
international scientists of the Chinese Academy of Sciences
(2013T2J0057). This research has used NASA's Astrophysical Data System (ADS).
\end{acknowledgements}

\end{document}